\documentclass[dvips]{acta}
\usepackage{supertabular,lscape,epsfig}
\usepackage{amssymb}
\usepackage{amsmath}
\usepackage{wasysym}

\SetPages{1}{0}

\SetVol{0}{2018}

\begin{document}

\begin{Titlepage}
\Title{Planet-star interactions with precise transit timing. I. The refined orbital decay rate for WASP-12~b and initial constraints for HAT-P-23~b, KELT-1~b, KELT-16~b, WASP-33~b, and WASP-103~b}
\Author{G.~Maciejewski$^{1}$,  M.~Fern\'andez$^2$,  F.~Aceituno$^2$,  S.~Mart\'{\i}n-Ruiz$^2$,   J.~Ohlert$^{3,4}$,  D.~Dimitrov$^{5,6}$, K.~Szyszka$^{1}$,  C.~von~Essen$^7$,  M.~Mugrauer$^8$,  R.~Bischoff $^8$,  K.-U.~Michel$^8$,  M.~Mallonn$^9$,  M.~Stangret$^{1}$,  D.~Mo\'zdzierski$^{10}$}
{$^1$Centre for Astronomy, Faculty of Physics, Astronomy and Informatics,
         Nicolaus Copernicus University, Grudziadzka 5, 87-100 Toru\'n, Poland,
         e-mail: gmac@umk.pl\\
 $^2$Instituto de Astrof\'isica de Andaluc\'ia (IAA-CSIC), Glorieta de la Astronom\'ia 3, 18008 Granada, Spain\\
 $^3$Michael Adrian Observatorium, Astronomie Stiftung Trebur, 65428 Trebur, Germany\\
 $^4$University of Applied Sciences, Technische Hochschule Mittelhessen, 61169 Friedberg, Germany\\
 $^5$Institute of Astronomy and NAO, Bulgarian Academy of Science, 1784 Sofia, Bulgaria\\
 $^6$Department of Physics and Astronomy, Shumen University, 9700 Shumen, Bulgaria\\
 $^7$Stellar Astrophysics Centre, Department of Physics and Astronomy, Aarhus University, Ny Munkegade 120, DK-8000 Aarhus C,
Denmark\\
 $^8$Astrophysikalisches Institut und Universit\"ats-Sternwarte, Schillerg{\"a}sschen 2, 07745 Jena, Germany\\
 $^9$Leibniz-Institut f{\"u}r Astrophysik Potsdam, An der Sternwarte 16, D-14482 Potsdam, Germany\\
 $^{10}$Astronomical Institute, University of Wroc\l{}aw, Kopernika 11, 51-622 Wroc\l{}aw, Poland}

\Received{December 6, 2018}
\end{Titlepage}

\Abstract{Theoretical calculations and some indirect observations show that massive exoplanets on tight orbits must decay due to tidal dissipation within their host stars. This orbital evolution could be observationally accessible through precise transit timing over a course of decades. The rate of planetary in-spiralling may not only help us to understand some aspects of evolution of planetary systems, but also can be used as a probe of the stellar internal structure. In this paper we present results of transit timing campaigns organised for a carefully selected sample of hot Jupiter-like planets which were found to be the best candidates for detecting planet-star tidal interactions on the Northern hemisphere. Among them, there is the WASP-12 system which is the best candidate for possessing an in-falling giant exoplanet. Our new observations support the scenario of orbital decay of WASP-12~b and allow us to refine its rate. The derived tidal quality parameter of the host star $Q'_{*} = (1.82 \pm 0.32) \times 10^{5}$ is in agreement with theoretical predictions for subgiant stars. For the remaining systems -- HAT-P-23, KELT-1, KELT-16, WASP-33, and WASP-103 -- our transit timing data reveal no deviations from the constant-period models, hence constraints on the individual rates of orbital decay were placed. The tidal quality parameters of host stars in at least 4 systems -- HAT-P-23, KELT-1, WASP-33, and WASP-103 -- were found  to be greater than the value reported for WASP-12. This is in line with the finding that those hosts are main sequence stars, for which efficiency of tidal dissipation is predicted to be relatively weak.}{planet-star interactions -- stars: individual: HAT-P-23, KELT-1, KELT-16, WASP-12, WASP-33, WASP-103 -- planets and satellites: individual: HAT-P-23 b, KELT-1 b, KELT-16 b, WASP-12 b, WASP-33 b, WASP-103 b}

%%%%%%%%%%%%%%%%%%%%%%%%%%%%%%%%%%%%%%%%%%%%%%%%%%%%%%%%%%%%%%%%%%%%%%

\section{Introduction}

Since the first detections of massive planets on tight orbits around solar-like stars (Latham \etal 1989, Mayor \& Queloz 1995, Marcy \& Butler 1996) it has been possible to investigate planet-star tidal interactions outside the Solar Sytem. The mechanism of tides is well studied in the Earth-Moon-Sun system. The tidal force, which is raised by one body on the other, is proportional to the mass of the body rising the tide, and inversely proportional to the cube of the distance between both bodies. For some very hot Jupiter-like exoplanets (with orbital periods shorter than a couple of days), the orbital separation is as small as several stellar radii, making such systems great laboratories for studies of tidal effects. 

The tidal and rotational bulges deform figures of both the star and the planet. This departure from spherical symmetry entails that the quadrupole component of the gravitational field makes the planetary orbit to precess. Ragozzine \& Wolf (2009) found that apsidal precession for very hot Jupiters is driven mainly by the quadrupole of the planetary tidal bulge. Its rate depends on properties of the planetary interior. Calculations show that the precession period could be as short as a few decades for the most promising systems (e.g., Birkby \etal 2014). The precession motion could be detected using the method of precise transit timing if only the planetary orbit is non-circular. However, tidal interactions tend to circularise planetary orbits in relatively short timescales, so any non-zero eccentricity would need to be invoked and sustained by a certain process, such as gravitational perturbations induced by nearby low-mass planets. 

A bulk of hot Jupiters are far from tidal equilibrium. They are expected to be spiralling towards host stars because of the dissipative nature of tides, caused by the friction of the tidally induced fluid flow (Levrard \etal 2009). In such systems, the host star usually rotates slower than the planet orbits it. There is a phase lag in the tidal response that results in transferring the orbital angular momentum into the star -- the orbit shrinks and the star spins up. The dissipation of energy stored in the equilibrium tides is thought to occur in stellar zones where the viscosity is induced by the turbulent convective motions (Zahn 1966, Goldreich \& Nicholson 1977). The numerical simulations show, however, that the effective viscosity can be reduced in the systems with high tidal frequency (e.g., Ogilvie \& Lesur 2012). On the other hand, the tidal dissipation can be boosted by radiative damping of the dynamical tides that are produced near radiative-convective boundaries (Goldreich \& Nicholson 1989 and reference therein). 

The efficiency of tidal dissipation in the host star can be characterised with the dimensionless tidal quality parameter 
\begin{equation}
  Q'_{*} = \frac{3}{2}\frac{Q_{*}}{k_2}\, , \;
\end{equation}
(Goldreich \& Soter 1966), where $Q_{*}$ is the inverse of the phase lag between the tidal potential and the tidal bulge (or the ratio of energy stored in tidal distortion to energy dissipated in one tidal cycle), and $k_2$ is the second order tidal Love number which is a dimensionless measure of the density profile inside the star. The tidal quality parameter depends on the properties of the star such as the mass, structure, and rotation rate. A smaller value of $Q'_{*}$ translates into a stronger or more efficient tidal dissipation and vice versa. Theoretical studies of turbulent damping of equilibrium tides predict $Q_{*}$ of $10^8$--$10^9$ for main-sequence stars (Penev \& Sasselov 2011). The studies of binary stars in stellar clusters show that $Q'_{*}$ might be of order $10^6$ (Meibom \& Mathieu 2005, Ogilvie \& Lin 2007, Milliman \etal 2014). On the other hand, a statistical analysis of the destruction rate of hot Jupiters, which is based on the population of currently known planets, yields $Q'_{*}>10^7$ (Penev \etal 2012). Jackson \etal (2008) and Husnoo \etal (2012) obtained $Q'_{*} \sim 10^{6.5}$, while Hansen (2010) found $10^7 < Q'_{*} < 10^9$. An investigation of orbital parameters for a sample of 231 hot planets allowed Bonomo \etal (2017) to conclude that $Q'_{*}$ must exceed $10^6$--$10^7$. Penev \etal (2018) studied hot-Jupiter systems with known stellar rotation periods and found that the tidal quality parameter may also depend on the amplitude and frequency of the tidal excitation. The authors concluded that the value of $Q'_{*}$ might be in a wide range between $10^5$ and $10^7$ with the dependence on tidal frequency. This dependency could reconcile apparently inconsistent values of $Q'_{*}$ which are reported in the literature. 
 
Orbital decay can be detected through transit timing. For some planets, the cumulative shift in transit times may be of order 100 s after ten years if $Q'_{*}=10^6$ is assumed (Birkby \etal 2014, Essick \& Weinberg 2016). So far, the only candidate for a spiralling-in exoplanet is WASP-12~b (Hebb \etal 2009). The planet has a mass of $\sim 1.4$ $M_{Jup}$ (Jupiter mass) and bloated radius of $\sim 1.9$ $R_{Jup}$ (Jupiter radius). It orbits its F/G host star within 1.09 d. Maciejewski \etal (2016) employed the method of precise transit timing to detect the apparent shortening of the orbital period. Departure from a linear transit ephemeris by $\sim 5$ minutes was observed in the course of 8 years. This finding translates into the rate of orbital period shortening of $\sim2.6 \times 10^{-2}$ s~yr$^{-1}$, giving $Q'_{*} = 2.5 \times 10^{5}$. Although transit times were found to follow the quadratic ephemeris very well, there is still an alternative scenario in which the observed period shrinkage is de facto a part of a long-period cycle caused by either tidally induced orbital precession (Maciejewski \etal 2016, Patra \etal 2017) or dynamical interactions with a planetary companion (Maciejewski 2018). New transit timing observations are expected to help distinguish between both models.

Being motivated by the case of the WASP-12 system, we have initiated a systematic transit timing monitoring programme for a sample of hot giant exoplanets for which period shrinkage could be detected in the course of a decade. In this paper, we discuss sample selection criteria and present the first results obtained for the systems of the sample.
 
%%%%%%%%%%%%%%%%%%%%%%%%%%%%%%%%%%%%%%%%%%%%%%%%%%%%%%%%%%%%%%%%%%%%%%

\section{The sample}

\subsection{Selection criteria}

Our procedure of sample selection was based on the catalogue of transiting exoplanets which was provided by the Extrasolar Planets Encyclopaedia\footnote{http://exoplanet.eu} (Schneider \etal 2011). Using an on-line tool, a list of confirmed transiting exoplanets was extracted. With one-meter class telescopes, which are used by us for transit observations, it is possible to acquire photometric time series with a precision of $\sim1$ ppth (parts per thousand) of the normalised flux per minute for stars brighter than 12--13 mag. Such datasets allow us to determine individual mid-transit times with errors between 20 and 40 s, and when they are combined together, an averaged  timing precision down to or even below 10 s ($\approx 1 \sigma$) can be achieved in time scales of months. Therefore we considered only planets for which the predicted cumulative shift in transit times $T_{\rm{shift}}$ is greater than 30 s ($>3 \sigma$) after 10 years. To estimate $T_{\rm{shift}}$ for individual systems, we transformed the equation (20) of Goldreich \& Soter (1966) to derive
\begin{equation}
  T_{\rm{shift}} = - \frac{27}{4} \frac{\pi}{Q'_{*}} \left( \frac{M_{p}}{M_{*}} \right) {\left( \frac{a}{R_{*}} \right)}^{-5} \frac{1}{P_{\rm{orb}}}  {\left( 10 \: \rm{yr} \right)}^2\, , \;
\end{equation}
where $P_{\rm{orb}}$ is the orbital period, $M_{p}$ is the planet's mass, $M_{*}$ and $R_{*}$ are the host star's mass and radius, and $a$ is the semi-major axis of the planet's orbit. In our calculations, the canonical value of $Q'_{*} = 10^6$ was assumed. Since our one-meter class telescopes are located on the northern hemisphere, we considered only systems with declination greater than $0^{\circ}$. To meet the photometric quality criterion, we rejected systems with stars fainter than $\sim 13$ mag in the $V$ band and systems with transits shallower than 3 ppth. 

Six systems passed our criteria: HAT-P-23, KELT-1, KELT-16, WASP-12, WASP-33, and WASP-103. Their values of $T_{\rm{shift}}$ were found to be in a range of --640 (KELT-1~b) to --34 (WASP-33~b) s per 10 years. Since the uncertainty of the assumed $Q'_{*}$ dominates the error budget of $T_{\rm{shift}}$, we treat the calculated values as estimates without uncertainties. Figure~1 shows the location of the exoplanets of our sample in the diagram of $T_{\rm{shift}}$ vs. $P_{\rm{orb}}$.

% FIGURE Sample selection
\begin{figure}[thb]
\begin{center}
\includegraphics[width=0.7\textwidth]{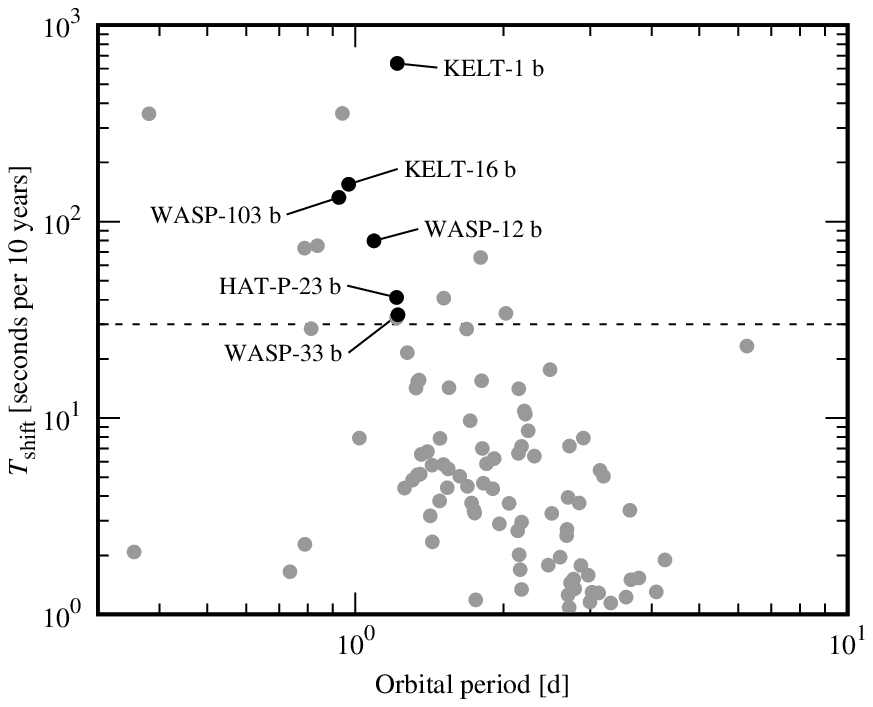}
\end{center}
\FigCap{Predicted mid-transit time shifts $T_{\rm{shift}}$ vs. orbital periods $P_{\rm{orb}}$ for all known transiting planets (grey dots) and 6 selected exoplanets of our sample (black dots). The dashed line marks the adopted threshold of 30 s per 10 years.}
\end{figure}

\subsection{Objects of the sample}

HAT-P-23~b is a 2.1 $M_{\rm{Jup}}$ planet orbiting an early G dwarf star within 29 h (Bakos \etal 2011). The radial velocity follow-ups show that its orbit is prograde and aligned (Moutou \etal 2011), and together with occultation observations (O'Rourke \etal 2014) show that the orbit is circular with a $2 \sigma$ upper limit for its eccentricity equal to 0.052 (Bonomo \etal 2017). The planet was initially found to have a radius of $R_{\rm{b}} \approx 1.4$ $R_{\rm{Jup}}$, so it was classified as a highly inflated transiting planet (Bakos \etal 2011). However, the follow-up transit observations by Ram\'on-Fox \& Sada (2013), Ciceri \etal (2015), and Sada \& Ram\'on-Fox (2016) do not confirm that finding, yielding a more compact planet with $R_{\rm{b}} \approx 1.2$ $R_{\rm{Jup}}$. One must note, however, that this discrepancy is produced by a smaller value of the redetermined stellar radius in a study of Ciceri \etal (2015) while the value of the planet-to-star radii ratio $k$ is reproduced, and by an apparently smaller value of $k$ reported by Ram\'on-Fox \& Sada (2013) and Sada \& Ram\'on-Fox (2016).

With a mass of $\sim 27$ $M_{\rm{Jup}}$, KELT-1~b was classified as a low-mass brown dwarf or a super-massive planet (Siverd \etal 2012). The boundary between planets and brown dwarfs is not clearly defined, ranging from $\sim 10$ $M_{\rm{Jup}}$ of the  formation-based definition (Schlaufman 2018), through $\sim 13$ $M_{\rm{Jup}}$ from a deuterium burning criterion (Burrows \etal 2001, Spiegel \etal 2011), to $25-30$ $M_{\rm{Jup}}$ resulting from the intersection of the mass distribution of sub-stellar objects (Udry 2010). KELT-1~b fulfils criteria of being a high-mass giant gaseous planet as defined by Hatzes \& Rauer (2015). The body orbits its mid-F star on a circular and aligned orbit within 29 h (Siverd \etal 2012, Beatty \etal 2014, Croll \etal 2015). With a radius of 1.1 $R_{\rm{Jup}}$, the planet was found to be significantly inflated when compared to standard evolutionary models without insolation (Siverd \etal 2012). Occultation observations indicate that the day side has an average brightness temperature of $\sim$3200~K (e.g., Beatty \etal 2017) and there is a weak or moderate redistribution of heat in the atmosphere (Beatty \etal 2014). Baluev \etal (2015) used amateur transit observations to refine a transit ephemeris and found that the orbital period is $\sim 0.6$ s shorter compared to that one in the discovery paper. 

KELT-16~b belongs to a group of exoplanets with orbital periods shorter than 1~d. It has a mass of $\sim$2.8 $M_{\rm{Jup}}$ and a radius of 1.4 $R_{\rm{Jup}}$, and needs only 23.5 h to orbit its F7 host star (Oberst \etal 2017). In this paper, we present the first follow-up observations acquired a number of cycles after the discovery of the system.

WASP-12~b is the only hot Jupiter planet for which long-term variations in transit times have been detected from ground observations (Maciejewski \etal 2016). At the time of discovery, it was found to be one of the most intensely irradiated planets (Hebb \etal 2009). The planet's proximity to the star results in a high equilibrium temperature of 2500 K, thus encouraging numerous studies on the properties of the planetary atmosphere. The optically opaque planet is tidally distorted and occupies a significant fraction of the Roche lobe (e.g.\ Budaj 2011). Observations indicate that the globe is  surrounded by a translucent exosphere which overfills the Roche lobe and escapes, forming a stable and translucent circumstellar cloud (Debrecht \etal 2018 and references therein).

The A5 star WASP-33 was found to harbour a $\sim$3.3 $M_{\rm{Jup}}$ and $\sim$1.5 $R_{\rm{Jup}}$ giant planet on a 29 h orbit (Christian \etal 2006, Kov\'acs \etal 2013). The fast rotation of the star precludes acquiring accurate radial velocity measurements, so the line-profile tomography during transit was used to confirm the planetary nature of WASP-33~b (Collier Cameron \etal 2010). The retrograde, nearly polar orbit was found to undergo nodal precession (Johnson \etal 2015) which is a manifestation of classical and general relativistic non-Keplerian orbital effects (Iorio 2011). The star was found to be a non-radial pulsator of the $\delta$ Scuti type (Collier Cameron \etal 2010, Herrero \etal 2011). Von Essen \etal (2014) identified 8 discrete pulsation components with amplitudes between 0.4 and 1.0 mmag. Occultation observations result in a high brightness temperature, suggesting that the redistribution of heat from the day side of the globe to the night side, as well as within the day side, is highly inefficient (Smith \etal 2011).

With a period as short as 22.2 h, WASP-103~b belongs to a group of ultra-short-period giant planets close to tidal disruption (Gillon \etal 2014). It has a mass of $\sim1.5$ $M_{\rm{Jup}}$ and radius of $\sim 1.5$ $R_{\rm{Jup}}$, and orbits an F8 dwarf. Southworth \etal (2015) acquired 17 follow-up transit light curves that allowed them to refine the transit ephemeris and to detect a variation of the planetary radius with optical wavelength. This effect was thought to be caused in part by Rayleigh scattering in the planetary atmosphere (Southworth \etal 2015, Southworth \& Evans 2016, Turner \etal 2017). However, the strong scattering slope is not supported by transmission spectroscopy observations (Lendl \etal 2017). The system parameters were revised by a global analysis of Delrez \etal (2018) in which new transit and occultation photometric time series were used together with literature data.

%%%%%%%%%%%%%%%%%%%%%%%%%%%%%%%%%%%%%%%%%%%%%%%%%%%%%%%%%%%%%%%%%%%%%%

\section{Observations and data reduction}

Photometric observations, which are reported in this paper, were acquired between March 2016 and November 2018. A bulk of the photometric time series were acquired with instruments which are listed below according to mirror diameter:
\begin{itemize}
 \item the 2.0 m Ritchey-Chr\'etien-Coud\'e telescope at the National Astronomical Observatory Rozhen (Bulgaria), equipped with a Roper Scientific VersArray 1300B (by December 2017) or ANDOR iKon-L 936 (from January 2018) CCD camera -- ROZ 2.0,
 \item the 1.5 m Ritchey-Chr\'etien Telescope at the Sierra Nevada Observatory (OSN, Spain) equipped with a Roper Scientific VersArray 2048B CCD camera -- OSN 1.5,
  \item the 1.2 m Cassegrain telescope at the Michael Adrian Observatory (Trebur, Germany), equipped with an SBIG STL-6303E CCD camera -- TRE 1.2,
  \item the 0.9 m Ritchey-Chr\'etien Telescope at the OSN, equipped with a Roper Scientific VersArray 2048B CCD camera -- OSN 0.9,
  \item the 0.6 m Cassegrain telescope at the Centre for Astronomy of the Nicolaus Copernicus University (Piwnice near Toru\'n, Poland), equipped with an SBIG STL-1001M (by June 2018) or FLI ML16803 (from August 2018) CCD camera -- PIW 0.6.
\end{itemize}
Moreover, some additional data were acquired with 
\begin{itemize}
 \item the 10.4 m Gran Telescopio Canarias at the Observatorio del Roque de los Muchachos (ORM, La Palma, Spain) and the long slit OSIRIS spectrograph -- GTC 10.4, 
 \item the 2.2 m reflector at the Calar Alto Astronomical Observatory (Spain) and the Calar Alto Faint Object Spectrograph (CAFOS) -- CAHA 2.2,
 \item the 2.0 m Liverpool Telescope (Steele \etal 2004) at the ORM and the RISE fast-readout camera (Steele \etal 2008) -- LT 2.0,
  \item the 0.9 m telescope at the University Observatory Jena (Germany) and the Schmidt Teleskop Kamera (STK) with an E2V CCD42-40 (grade 1) detector (Mugrauer \& Berthold 2010) -- JEN 0.9,
  \item the 0.6 m Cassegrain telescope at the Bia\l{}k\'ow Observatory of the Wroc\l{}aw University (Poland), equipped with an Andor CCD camera -- BIA 0.6.
\end{itemize}
To suppress flat-field inaccuracies, all telescopes were automatically or manually guided with a precision of a few arc seconds. The instrumental set-ups were usually defocused to avoid saturation for longer exposures that reduced dead-time needed for a CCD read-out. Observations usually started 1--2 hours before an expected transit ingress and lasted additional 1--2 hours after the egress. Those out-of-transit portions of the data were used to monitor trends in photometry that were caused by differential atmospheric extinction and instrumental effects. These requirements were not always met, mainly due to technical or weather reasons. A bulk of the time series were acquired in white light or through a blue blocking ($\lambda < 500$ nm) long-pass filter (LP500). This approach allowed us to increase the instrumental efficiency, resulting in a higher signal-to-noise ratio which is crucial for precise transit purposes. Some light curves were acquired through Cousins $R$ filters in which the instruments were found to have the highest efficiency. One transit of WASP-12~b was observed in the Johnson $B$ band, and another one through a wide $V$+$R$ filter, optimised for transit observations of Sun-like stars. For WASP-33, three light curves recorded in the Johnson $V$ band to avoid significant defocusing and hence blending. Since the time survey plays a pivotal role for timing purposes, the timestamps were synchronised to UTC with at least sub-second accuracy via GPS or Network Time Protocol.
 
In total, we acquired 82 photometric time series: 16 for 13 transits of HAT-P-23~b (one transit was observed simultaneously with two instruments and another one with three), 9 transit light curves for KELT-1~b, 11 for KELT-16~b, 22 for 19 transits of WASP-12~b, 11 for WASP-33, and 13 for 12 transits of WASP-103~b. Details on individual runs are listed in Table~1, sorted by target names and then by dates.

\MakeTable{r l l c l c c c c}{12.5cm}{Summary for the new transit light curves acquired.}
{\hline
ID & Date  & Telescope & Filter  &  Airmass change  & $N_{\rm{obs}}$ & $t_{\rm{e}}$ (s) & $\Gamma$ & $pnr$\\
 \hline
\multicolumn{9}{c}{HAT-P-23 b}\\ 
1 & 2016 Jul 02 & OSN 1.5  & $R_{\rm{C}}$ & $1.53\rightarrow1.07\rightarrow1.15$ & 395 & 40 & 1.33 & 0.93 \\
2 & 2016 Aug 21 & OSN 0.9  & $R_{\rm{C}}$ & $1.07\rightarrow1.72$ & 227 & 60 & 0.91 & 1.57 \\
3 &                        & OSN 1.5  & $R_{\rm{C}}$ & $1.07\rightarrow1.72$ & 359 & 20 & 2.73 & 1.13 \\
4 & 2016 Aug 26 & PIW 0.6  & none & $1.36\rightarrow1.24\rightarrow1.39$ & 721 & 15 & 2.99 & 1.50 \\
5 & 2016 Sep 01 & OSN 1.5  & $R_{\rm{C}}$ & $1.18\rightarrow1.07\rightarrow1.28$ & 255 & 40 & 1.33 & 1.11 \\
6 &                       & TRE 1.2  & none & $1.26\rightarrow1.19\rightarrow1.47$ & 274 & 45 & 1.13 & 0.86  \\
7 &                       & PIW 0.6  & none & $1.28\rightarrow1.24\rightarrow1.69$ & 459 & 25 & 2.00 & 1.73 \\
8 & 2016 Sep 07 & TRE 1.2  & none & $1.20\rightarrow2.32$ & 288 & 45 & 1.13 & 1.48 \\
9 & 2017 Jul 23 & OSN 1.5  & $R_{\rm{C}}$ & $1.34\rightarrow1.07\rightarrow1.11$ & 319 & 40 & 1.33 & 0.96  \\
10 & 2017 Jul 30 & OSN 1.5  & $R_{\rm{C}}$ & $1.10\rightarrow1.07\rightarrow1.39$ & 650 & 20 & 2.73 & 0.93 \\
11 & 2017 Aug 16 & OSN 1.5  & $R_{\rm{C}}$ & $1.08\rightarrow1.07\rightarrow1.48$ & 654 & 20 & 2.73 & 1.03  \\
12 & 2017 Oct 16 & TRE 1.2  & none & $1.19\rightarrow2.04$ & 226 & 55 & 0.95 & 1.05 \\
13 & 2018 Jul 17 & TRE 1.2  & none & $1.47\rightarrow1.19\rightarrow1.22$ & 134 & 60 & 0.88 & 1.46 \\
14 & 2018 Aug 03 & TRE 1.2  & none & $1.36\rightarrow1.19\rightarrow1.40$ & 182 & 80 & 0.68 & 1.13 \\
15 & 2018 Aug 08 & ROZ 2.0  & none & $1.43\rightarrow1.13$ & 210 & 25 & 1.93 & 0.74 \\
16 & 2018 Sep 06 & ROZ 2.0  & none & $1.11\rightarrow1.72$ & 367 & 25 & 1.94 & 0.67 \\
\multicolumn{9}{c}{KELT-1 b}\\ 
1 & 2017 Jul 25 & OSN 0.9  & none & $1.61\rightarrow1.00$ & 980 & 12 & 3.99 & 0.77 \\
2 & 2017 Aug 15 & PIW 0.6  & BB500 & $1.60\rightarrow1.03\rightarrow1.04$ & 1044 & 15 & 2.99 & 1.68 \\
3 & 2017 Sep 19 & OSN 0.9  & none & $1.03\rightarrow1.00\rightarrow1.32$ & 707 & 20 & 2.61 & 0.71 \\
4 & 2017 Sep 23 & OSN 0.9  & none & $1.34\rightarrow1.00\rightarrow1.04$ & 747 & 20 & 2.61 & 0.77 \\
5 & 2017 Sep 30 & PIW 0.6  & BB500 & $1.05\rightarrow1.03\rightarrow1.58$ & 769 & 15 & 2.99 & 1.21 \\
6 & 2017 Nov 23 & PIW 0.6  & BB500 & $1.10\rightarrow1.03\rightarrow1.30$ & 470 & 25 & 2.00 & 1.04 \\
7 & 2018 Jan 07 & PIW 0.6  & BB500 & $1.07\rightarrow2.13$ & 814 & 15 & 2.99 & 1.35 \\
8 & 2018 Sep 05 & PIW 0.6  & BB500 & $1.53\rightarrow1.03\rightarrow1.04$ & 490 & 30 & 1.58 & 1.33 \\
9 & 2018 Nov 05 & JEN 0.9  & none & $1.11\rightarrow1.02\rightarrow1.05$ & 210 & 45 & 1.05 & 1.29 \\
\multicolumn{9}{c}{KELT-16 b}\\ 
1 & 2016 Nov 21 & PIW 0.6  & BB500 & $1.13\rightarrow2.09$ & 321 & 35 & 1.50 & 2.11 \\
2 & 2017 Jun 22 & ROZ 2.0  & $R_{\rm{C}}$ & $1.25\rightarrow1.02$ & 258 & 25 & 1.36 & 0.94 \\
3 & 2017 Jul 23 & OSN 0.9  & none & $1.25\rightarrow1.00\rightarrow1.07$ & 711 & 20 & 2.61 & 0.90 \\
4 & 2017 Aug 23 & TRE 1.2  & none & $1.05\rightarrow1.68$ & 409 & 30 & 1.58 & 0.84 \\
5 & 2017 Sep 24 & OSN 0.9  & none & $1.01\rightarrow1.00\rightarrow1.76$ & 417 & 40 & 1.39 & 0.95 \\
6 & 2017 Sep 25 & OSN 0.9  & none & $1.02\rightarrow1.00\rightarrow1.49$ & 394 & 40 & 1.39 & 1.08 \\
7 & 2017 Sep 29 & BIA 0.6  & $R_{\rm{C}}$ & $1.11\rightarrow1.06\rightarrow1.26$ & 201 & 50 & 0.80 & 1.79 \\
8 & 2018 Aug 03 & PIW 0.6  & BB500 & $1.20 \rightarrow 1.07 \rightarrow 1.20$ & 473 & 25 & 1.82 & 1.42\\
9 & 2018 Sep 03 & PIW 0.6  & BB500 & $1.12 \rightarrow 1.07 \rightarrow 1.67$ & 342 & 52 & 1.02& 2.18\\
10 & 2018 Sep 06 & PIW 0.6  & BB500 & $1.14 \rightarrow 1.07 \rightarrow 1.20$ & 380 & 30 & 1.58 & 1.49\\
11 & 2018 Oct 09 & TRE 1.2  & none & $1.09 \rightarrow 1.05 \rightarrow 1.25$ & 250 & 50 & 1.03 & 0.71\\
\hline
\multicolumn{9}{l}{Date is given for the middle of the transit in UT. $N_{\rm{obs}}$ is the number of useful scientific exposures.}\\ 
\multicolumn{9}{l}{$t_{\rm{e}}$ is the exposure time used. $\Gamma$ is the median number of exposures per minute. $pnr$ is the}\\ 
\multicolumn{9}{l}{photometric scatter in parts per thousand of the normalised flux per minute of  observation, see}\\ 
\multicolumn{9}{l}{Fulton \etal (2011).}\\ 
}

\addtocounter{table}{-1}

\MakeTable{r l l c l c c c c l}{12.5cm}{Continued.}
{\hline
ID & Date  & Telescope & Filter  &  Airmass change  & $N_{\rm{obs}}$ & $t_{\rm{e}}$ (s) & $\Gamma$ & $pnr$ \\
 \hline
\multicolumn{9}{c}{WASP-12 b}\\ 
1 & 2016 Mar 03 & OSN 1.5  & $R_{\rm{C}}$ & $1.01 \rightarrow 2.15$ & 605 & 30 & 1.82 & 0.70  \\
2 & 2016 Oct 30 & OSN 1.5  & $R_{\rm{C}}$ & $2.12 \rightarrow 1.02$ & 413 & 30 & 1.71 & 0.94  \\
3 & 2016 Nov 11 & OSN 1.5  & $R_{\rm{C}}$ & $1.27 \rightarrow 1.01 \rightarrow 1.02$ & 274 & 40 & 1.33 & 0.73  \\
4 & 2016 Dec 03 & TRE 1.2  & none & $1.68 \rightarrow 1.07 \rightarrow 1.69$ & 355 & 50 & 1.03 & 0.80 \\
5 & 2016 Dec 05 & TRE 1.2  & none & $1.13 \rightarrow 1.07 \rightarrow 1.51$ & 344 & 50 & 1.03 & 0.62  \\
6 & 2017 Jan 18 & TRE 1.2  & none & $1.76 \rightarrow 1.07 \rightarrow 1.10$ & 351 & 50 & 1.03 & 0.93 \\
7 & 2017 Jan 19 & TRE 1.2  & none & $1.26 \rightarrow 1.07 \rightarrow 1.37$ & 378 & 50 & 1.03 & 0.79 \\
8 & 2017 Feb 02 & OSN 1.5  & $R_{\rm{C}}$ & $1.01 \rightarrow 1.49$ & 474 & 30 & 1.88 & 0.78 \\
9 &                      & OSN 0.9  & none & $1.01 \rightarrow 1.53$ & 466 & 30 & 1.82 & 0.92 \\
10 & 2017 Feb 24 & TRE 1.2  & none & $1.07 \rightarrow 2.00$ & 272 & 50 & 1.03 & 0.86 \\
11 &                      & PIW 0.6  & BB500 & $1.09 \rightarrow 1.88$ & 430 & 25 & 2.00 & 1.17  \\
12 & 2017 Feb 27 & LT 2.0  & $V$+$R$ & $1.00 \rightarrow 2.07$ & 3116 & 5 & 11.97 & 0.74  \\
13 & 2017 Sep 30 & CAHA 2.2  & $B$ & $2.23 \rightarrow 1.03$ & 217 & 40 & 0.92 & 1.24 \\
14 & 2017 Oct 24 & OSN 1.5  & none & $1.44 \rightarrow 1.01 \rightarrow 1.02$ & 614 & 20 & 2.37 & 0.59 \\
15 & 2017 Nov 16 & OSN 1.5  & $R_{\rm{C}}$ & $1.61 \rightarrow 1.01 \rightarrow 1.08$ & 942 & 20 & 2.72 & 0.76 \\
16 & 2017 Nov 17 & OSN 1.5  & $R_{\rm{C}}$ & $1.15 \rightarrow 1.01 \rightarrow 1.26$ & 842 & 20 & 2.72 & 0.73 \\
17 & 2018 Jan 13 & PIW 0.6  & BB500 & $1.11 \rightarrow 1.09 \rightarrow 1.62$ & 621 & 25 & 2.00 & 1.22  \\
18 & 2018 Feb 05 & PIW 0.6  & BB500 & $1.20 \rightarrow 1.09 \rightarrow 1.78$ & 689 & 25 & 2.00 & 1.58  \\
19 &                      & TRE 1.2  & none & $1.13 \rightarrow 1.07$ & 158 & 50 & 1.03 & 0.98 \\
20 & 2018 Feb 07 & TRE 1.2  & none & $ 1.07 \rightarrow$ 2.28 & 278 & 55 & 0.95 & 0.97 \\
21 & 2018 Feb 16 & PIW 0.6  & BB500 & $1.21 \rightarrow 1.09 \rightarrow 1.34$ & 651 & 25 & 2.00 & 1.21 \\
22 & 2018 Feb 28  & PIW 0.6  & BB500 & $1.10 \rightarrow 1.09 \rightarrow 1.38$ & 688 & 15 & 3.00 & 1.62 \\
\multicolumn{9}{c}{WASP-33 b}\\ 
1 & 2014 Aug 09 & GTC 10.4  & none & $2.07\rightarrow1.02$ & 587 & 1 & 2.40 & 0.86 \\
2 & 2014 Aug 31 & GTC 10.4  & none & $2.00\rightarrow1.02$ & 827 & 0.4 & 3.16 & 0.82 \\
3 & 2016 Dec 30 & PIW 0.6  & BB500 & $1.15\rightarrow1.04\rightarrow1.27$ & 1039 & 15 & 2.99 & 1.54 \\
4 & 2017 Jan 10 & PIW 0.6  & BB500 & $1.09\rightarrow1.04\rightarrow1.60$ & 1568 & 10 & 3.99 & 1.29 \\
5 & 2017 Sep 19 & PIW 0.6  & BB500 & $1.33\rightarrow1.04\rightarrow1.10$ & 945 & 15 & 2.99 & 1.26 \\
6 & 2017 Sep 30 & BIA 0.6  & $V$ & $1.08\rightarrow1.03\rightarrow1.19$ & 442 & 10 & 1.71 & 2.02 \\
7 & 2017 Nov 17 & PIW 0.6  & BB500 & $1.71\rightarrow1.04\rightarrow1.15$ & 1363 & 15 & 2.99 & 0.99 \\
8 & 2018 Sep 20 & PIW 0.6  & BB500 & $1.65\rightarrow1.04\rightarrow1.11$ & 1589 & 13 & 3.75 & 1.21 \\
9 & 2018 Oct 12 & PIW 0.6  & BB500 & $1.33\rightarrow1.04\rightarrow1.28$ & 1739 & 12 & 4.01& 1.09 \\
10 & 2018 Nov 07 & JEN 0.9  & $V$ & $1.68\rightarrow1.05$ & 388 & 25 & 1.62 & 1.59 \\
11 & 2018 Nov 13 & JEN 0.9  & $V$ & $1.08\rightarrow1.03\rightarrow1.10$ & 351 & 25 & 1.62 & 0.78 \\
\multicolumn{9}{c}{WASP-103 b}\\ 
1 & 2016 May 07 & PIW 0.6  & none & $2.26\rightarrow1.44\rightarrow1.49$ & 569 & 25 & 2.00 & 2.39 \\
2 & 2016 May 08 & PIW 0.6  & none & $2.21\rightarrow1.44\rightarrow1.49$ & 480 & 25 & 2.00 & 2.71 \\
3 & 2016 May 09 & PIW 0.6  & none & $2.51\rightarrow1.44$ & 512 & 25 & 2.00 & 2.64 \\
\hline
\multicolumn{9}{l}{Date is given for the middle of the transit in UT. $N_{\rm{obs}}$ is the number of useful scientific exposures.}\\ 
\multicolumn{9}{l}{$t_{\rm{e}}$ is the exposure time used. $\Gamma$ is the median number of exposures per minute. $pnr$ is the}\\ 
\multicolumn{9}{l}{photometric scatter in parts per thousand of the normalised flux per minute of  observation, see}\\ 
\multicolumn{9}{l}{Fulton \etal (2011).}\\ 
}

\addtocounter{table}{-1}

\MakeTable{r l l c l c c c c l}{12.5cm}{Continued.}
{\hline
ID & Date  & Telescope & Filter  &  Airmass change  & $N_{\rm{obs}}$ & $t_{\rm{e}}$ (s) & $\Gamma$ & $pnr$ \\
 \hline
\multicolumn{9}{c}{WASP-103 b -- continued}\\ 
4 & 2017 May 04 & OSN 0.9  & $R_{\rm{C}}$ & $1.24\rightarrow1.15\rightarrow1.25$ & 258 & 40 & 1.39 & 1.74 \\
5 & 2017 May 16 & OSN 0.9  & $R_{\rm{C}}$ & $1.84\rightarrow1.15\rightarrow1.22$ & 393 & 40 & 1.39 & 1.69 \\
6 & 2017 Jun 23 & OSN 0.9  & none & $1.24\rightarrow1.15\rightarrow2.18$ & 413 & 40 & 1.39 & 1.36 \\
7 & 2018 May 11 & OSN 1.5  & none & $1.47\rightarrow1.15\rightarrow1.44$ & 854 & 20 & 2.72 & 0.85 \\
8 & 2018 May 12 & OSN 1.5  & none & $1.92\rightarrow1.15\rightarrow1.35$ & 937 & 20 & 2.73 & 0.84 \\
9 & 2018 Jun 06 & OSN 1.5  & $R_{\rm{C}}$ & $1.45\rightarrow1.15\rightarrow1.61$ & 557 & 30 & 1.88 & 1.01 \\
10 & 2018 Jun 06 & TRE 1.2  & none & $1.54\rightarrow1.36\rightarrow1.81$ & 280 & 40 & 1.13 & 1.76 \\
11 & 2018 Jun 18 & OSN 1.5  & $R_{\rm{C}}$ & $1.31\rightarrow1.15\rightarrow1.56$ & 655 & 25 & 2.23 & 1.05 \\
12 & 2018 Jun 19 & ROZ 2.0  & none & $1.22\rightarrow1.97$ & 207 & 45 & 1.15 & 0.95 \\
13 & 2018 Aug 09 & ROZ 2.0  & none & $1.24\rightarrow1.95$ & 284 & 30 & 1.71 & 1.11 \\
\hline
\multicolumn{9}{l}{Date is given for the middle of the transit in UT. $N_{\rm{obs}}$ is the number of useful scientific exposures.}\\ 
\multicolumn{9}{l}{$t_{\rm{e}}$ is the exposure time used. $\Gamma$ is the median number of exposures per minute. $pnr$ is the}\\ 
\multicolumn{9}{l}{photometric scatter in parts per thousand of the normalised flux per minute of  observation, see}\\ 
\multicolumn{9}{l}{Fulton \etal (2011).}\\ 
}

Data reduction and preliminary analysis of all observations, except those from GTC 10.4, were performed with AstroImageJ (AIJ, Collins \etal 2017). A standard calibration procedure, which included subtracting a median bias or dark frame and dividing by a median sky flat, was applied to science frames. Timestamps were converted to barycentric Julian dates in barycentric dynamical time $(\rm{BJD_{TDB}})$ using a built-in converter. Fluxes were derived with the differential aperture photometry method against an optimised set of comparison stars which were available in the field of view. The radius of the photometric aperture was related to the full width at half maximum of the averaged stellar profile, multiplied by a factor which was optimised for individual light curves. The value of this factor was usually in the range of 0.8--1.3, mainly depending on the scale of defocusing. If justified by the improvement in the goodness of fit (Collins \etal 2017), simultaneous de-trending against the airmass, position on the matrix, time, and seeing was applied along with a trial transit model. In the final light curves, fluxes were normalised to unity outside the transits. 

Two white light curves, acquired for WASP-33~b's transits with GTC 10.4, were generated from series of low-resolution spectra used for exoplanetary atmospheric studies. The data reduction and light curves extraction were performed with IRAF following the procedure described in details in von Essen \etal (2018).

The final light curves are available at the project web page\\
\centerline{http://www.home.umk.pl/\~{}gmac/TTV}
and via CDS. 

%%%%%%%%%%%%%%%%%%%%%%%%%%%%%%%%%%%%%%%%%%%%%%%%%%%%%%%%%%%%%%%%%%%%%%

\section{Data analysis}

The method of data analysis was similar for all 6 exoplanets of our sample. Some deviations from the adopted scheme were required for the WASP-12 and WASP-33 systems; they are described in subsections of Section 5. In the first step, the new light curves were modelled with the  Transit Analysis Package (TAP, Gazak \etal 2012) to redetermine systemic parameters and determine individual mid-transit times $T_{\rm{mid}}$. A transit phenomenon is coded with the model of Mandel \& Agol (2002) which is parametrised with the orbital inclination $i_{\rm{orb}}$, the semi-major axis scaled in star radii $a/R_*$, and the ratio of planet to star radii $R_p/R_*$.  Circular orbits were assumed for all planets of our sample. The limb darkening (LD) law in the quadratic form (Kopal 1950) was used to model the apparent emission across the stellar disk. Gaussian priors were set upon the linear and quadratic LD coefficients using values bi-linearly interpolated from tables of Claret \& Bloemen (2011) with formal uncertainties of 0.1 as the centre and width of the priors. In trial fitting runs, LD coefficients were kept free, and their values were found to agree with the theoretical predictions within 2--3 $\sigma$. This finding justified our final approach. For white-light data, the theoretical LD coefficients were calculated by averaging those of $B$, $V$, $R$, and $I$ bands. A similar approach was applied to the LP500 and $V$+$R$ data by averaging $V$, $R$, and $I$ and just $V$ and $R$ coefficients, respectively.

For each system, $i_{\rm{orb}}$, $a/R_*$, and $R_p/R_*$ were linked together for all new light curves. Mid-transit times were determined for each observed epoch, so if multiple photometric time series were available, their $T_{\rm{mid}}$'s were linked together. LD coefficients were linked for the data acquired in the same bands.

The best-fit solutions were found in a result of a Markov Chain Monte Carlo (MCMC) random walk process based on the Metropolis-Hastings algorithm and a Gibbs sampler. A time-correlated (red) noise was determined with the wavelet-based technique (Carter \& Winn 2009). The priors were taken from the most recent papers for the individual systems. In the final run, 10 MCMC chains with $10^6$ steps each were calculated. For each chain, the first 10\% of trials were rejected in order to compensate for the impact of the initial values. The median value of marginalised posteriori probability distributions were used to determine the best-fit parameters. Their 1-$\sigma$ uncertainties were derived as 15.9 and 84.1 percentile values of the cumulative distributions. 

To obtain a complete set of mid-transit times determined in a homogenous way, the similar procedure was applied to light curves available in the literature. In the light curve modelling procedure, TAP was additionally allowed to search for possible linear trends for each light curve in order to account for de-trending imperfections and to include associated uncertainties in the total error budget of the fit. Only redetermined mid-transit times with their uncertainties were considered in further studies.

The sets of mid-transit times were used to refine the linear transit ephemerides in the form 
\begin{equation}
     T_{\rm{mid}}(E) = T_0 + P_{\rm{orb}} \times E \, , \;
\end{equation}
where $E$ is a transit number from the reference epoch $T_0$ given in the discovery paper. The final values of $T_0$ and $P_{\rm{orb}}$ and their uncertainties were derived from the posterior probability distributions of those parameters generated with the MCMC algorithm. We employed 100 chains, each of which was $10^4$ steps long after discarding the first 1000 trials. The median value and 15.9 and 84.1 percentile values of each cumulative distribution were adopted as the best-fit parameter and its uncertainties, respectively. 

To place a constraint on the rate of the orbital decay, and hence on $Q'_*$, a quadratic ephemeris in the form
\begin{equation}
  T_{\rm{mid}}= T_0 + P_{\rm{orb}} \times E + \frac{1}{2} \frac{d P_{\rm{orb}}}{d E} \times E^2 \, , \;
\end{equation}
where $\frac{d P_{\rm{orb}}}{d E}$ is the change in the orbital period between succeeding transits, was tried. Again, the MCMC method with configuration parameters as for the linear case described above, was used to determine the best-fit parameters. The value of $Q'_*$ was calculated after rearranging eq. (2) to the form
\begin{equation}
  Q'_* = - \frac{27}{2}\pi \left( \frac{M_{p}}{M_{*}}\right) {\left( \frac{a}{R_{*}} \right)}^{-5} {\left( \frac{d P_{\rm{orb}}}{d E} \right)}^{-1} P_{\rm{orb}} \, . \;
\end{equation}
If no orbital decay was detected and $\frac{d P_{\rm{orb}}}{d E}$ was found to be indistinguishable from 0 within 2$\sigma$, the lower constraint on $Q'_*$ at the 95\% confidence level was placed from the 5th percentile of the posterior probability distribution of $\frac{d P_{\rm{orb}}}{d E}$.

\section{Results}

The redetermined transit parameters, which were obtained for the investigated systems, are collected in Table~2. The literature values are also given for comparison purposes. New mid-transit times together with the redetermined ones from previous studies are listed in Tables~3--8 in the Appendix. The new light curves together with the timing residuals from the best-fit ephemerides are plotted in Figures~2--8. Below we discuss the results for the individual planets.     

\MakeTable{ l c c c l c}{12.5cm}{Transit parameters for the investigated systems.}
{\hline
 Source & $i_{\rm{orb}}$ ($^{\circ}$) & $a/R_*$ & $R_p/R_*$ & \multicolumn{1}{c}{$P_{\rm{orb}}$ (d)} & $T_0$ (BJD$_{\rm{TDB}}$)\\
 & & & & & $+2450000$ \\
\hline
\multicolumn{6}{c}{HAT-P-23~b} \\
 ThisP & $85.23^{+0.54}_{-0.48}$  & $4.465^{+0.062}_{-0.059}$ & $0.11612^{+0.00058}_{-0.00060}$ & $1.212886400^{(+71)}_{(-73)}$ & 4852.26534(15)\\
 Ba11 & $85.1^{+1.5}_{-1.5}$  & $4.14^{+0.23}_{-0.23}$ & $0.1169^{+0.0012}_{-0.0012}$ & 1.212884(2) & 4852.26538(18)\\
 Ci15 & $85.74^{+0.95}_{-0.95}$  & $4.54^{+0.09}_{-0.09}$ & $0.11616^{+0.00081}_{-0.00081}$ & 1.21288287(17) & 4852.26599(20)\\
 Sa16 & $87.4^{+1.7}_{-2.0}$  & $4.26^{+0.13}_{-0.14}$ & $0.1113^{+0.0010}_{-0.0009}$ & 1.2128867(2) & 4852.26548(17)\\ 
\multicolumn{6}{c}{~} \\
\multicolumn{6}{c}{KELT 1~b} \\
 ThisP & $85.3^{+2.9}_{-2.6}$  & $3.55^{+0.12}_{-0.18}$ & $0.0762^{+0.0012}_{-0.0010}$ & $1.21749397(21)$ & 5909.29289(28)\\
 Si12 & $87.6^{+1.4}_{-1.9}$  & $3.619^{+0.055}_{-0.087}$ & $0.07806^{+0.00061}_{-0.00058}$ & 1.217501(18) & 5909.29280(23)\\
 Ba15 & $-$  & $-$ & $0.0783^{+0.0014}_{-0.0014}$ & 1.21749448(80) & 5909.29297(19)\\
\multicolumn{6}{c}{~} \\
\multicolumn{6}{c}{KELT 16~b} \\
 ThisP & $84.5^{+2.0}_{-1.4}$  & $3.238^{+0.084}_{-0.075}$ & $0.1076^{+0.0010}_{-0.0010}$ & $0.96899320(29)$ & $7247.24774(24)$\\
 Ob17 & $84.4^{+3.0}_{-2.3}$  & $3.23^{+0.12}_{-0.13}$ & $0.1070^{+0.0013}_{-0.0012}$ & 0.9689951(24) & 7247.24791(19)\\
\multicolumn{6}{c}{~} \\
\multicolumn{6}{c}{WASP-12~b} \\
 ThisP & $82.87^{+0.44}_{-0.41}$  & $3.026^{+0.022}_{-0.022}$ & $0.11753^{+0.00033}_{-0.00034}$ & $1.09142172(15)$ & 4508.97694(13)\\
 Ma13 & $82.96^{+0.50}_{-0.44}$  & $3.033^{+0.022}_{-0.021}$ & $0.1173^{+0.0005}_{-0.0005}$ & 1.0914209(2) & 4508.97718(22)\\
\multicolumn{6}{c}{~} \\
\multicolumn{6}{c}{WASP-33~b} \\
 ThisP & $88.5^{+1.1}_{-1.4}$  & $3.65^{+0.02}_{-0.04}$ & $0.10527^{+0.00061}_{-0.00059}$ & $1.219870897(79)$ & 4163.22449(16)\\
 Ko13 & $86.2^{+0.2}_{-0.2}$  & $3.69^{+0.01}_{-0.01}$ & $0.1143^{+0.0002}_{-0.0002}$ & $-$ & $-$\\
 vE14 & $87.90^{+0.93}_{-0.93}$  & $3.68^{+0.03}_{-0.03}$ & $0.1046^{+0.0006}_{-0.0006}$ & $1.2198675(11)$ & 4163.2282(3)\\
 Zh18 & $87.6^{+1.4}_{-1.2}$  & $3.65^{+0.04}_{-0.05}$ & $0.1055^{+0.0011}_{-0.0011}$ & $1.21987089(15)$ & 4163.22367(22)\\
\multicolumn{6}{c}{~} \\
\multicolumn{6}{c}{WASP-103~b} \\
 ThisP & $87.9^{+1.4}_{-1.7}$  & $2.996^{+0.018}_{-0.033}$ & $0.1120^{+0.0007}_{-0.0007}$ & $0.925545333^{(+91)}_{(-90)}$ & $6459.59934(10)$\\
 Gi14 & $86.3^{+2.7}_{-2.7}$  & $2.978^{+0.050}_{-0.096}$ & $0.1093^{+0.0019}_{-0.0017}$ & $0.925542(19)$ & 6459.59957(75)\\
 So15 & $87.3^{+1.2}_{-1.2}$  & $2.999^{+0.031}_{-0.031}$ & $0.1127^{+0.0009}_{-0.0009}$ & $0.9255456(13)$ & 6459.599386(55)\\
 De18 & $88.8^{+0.8}_{-1.1}$  & $3.010^{+0.008}_{-0.013}$ & $0.1150^{+0.0020}_{-0.0015}$ & $0.92554517(58)$ & 6459.599543(63)\\
\hline
\multicolumn{6}{l}{ThisP -- this paper, Ba11 -- Bakos \etal (2011), Ci15 -- Ciceri \etal (2015), Sa16 -- Sada \& Ra-} \\
\multicolumn{6}{l}{m\'on-Fox (2016), Si12 -- Siverd \etal (2012), Ba15 -- Baluev \etal (2015), Ob17 -- Oberst \etal} \\
\multicolumn{6}{l}{(2017), model YY Circular, Ma13 -- Maciejewski \etal (2013), Ko13 -- Kov\'acs \etal (2013),} \\
\multicolumn{6}{l}{vE14 -- von Essen \etal (2014), Zh18 -- Zhang \etal (2018), Gi14 -- Gillon \etal (2014), So15 --} \\
\multicolumn{6}{l}{Southworth \etal (2015), De18 -- Delrez \etal (2018).} \\
}

\subsection{HAT-P-23 b}

Our determinations of the transit parameters are based on 16 light curves of millimag and sub-millimag precision (Fig.~2). While compared to the previous studies, they are found to be consistent within 1--2$\sigma$. We note however that our study gives the most accurate results reported so far. The smaller values of $a/R_*$ that are reported by Bakos \etal (2011) and Sada \& Ram\'on-Fox (2016) are caused by adopting non-zero eccentricity by those authors. Since the recent study by Bonomo \etal (2017) shows that HAT-P-23~b's eccentricity remains undistinguishable from zero, adopting a circular orbit is justified. In Table~3, we do not list the values reported by Ram\'on-Fox \& Sada (2013) because they are obsoleted by results of Sada \& Ram\'on-Fox (2016). 

% FIGURE HAT-P-23
\begin{figure}[thb]
\begin{center}
\includegraphics[width=1.0\textwidth]{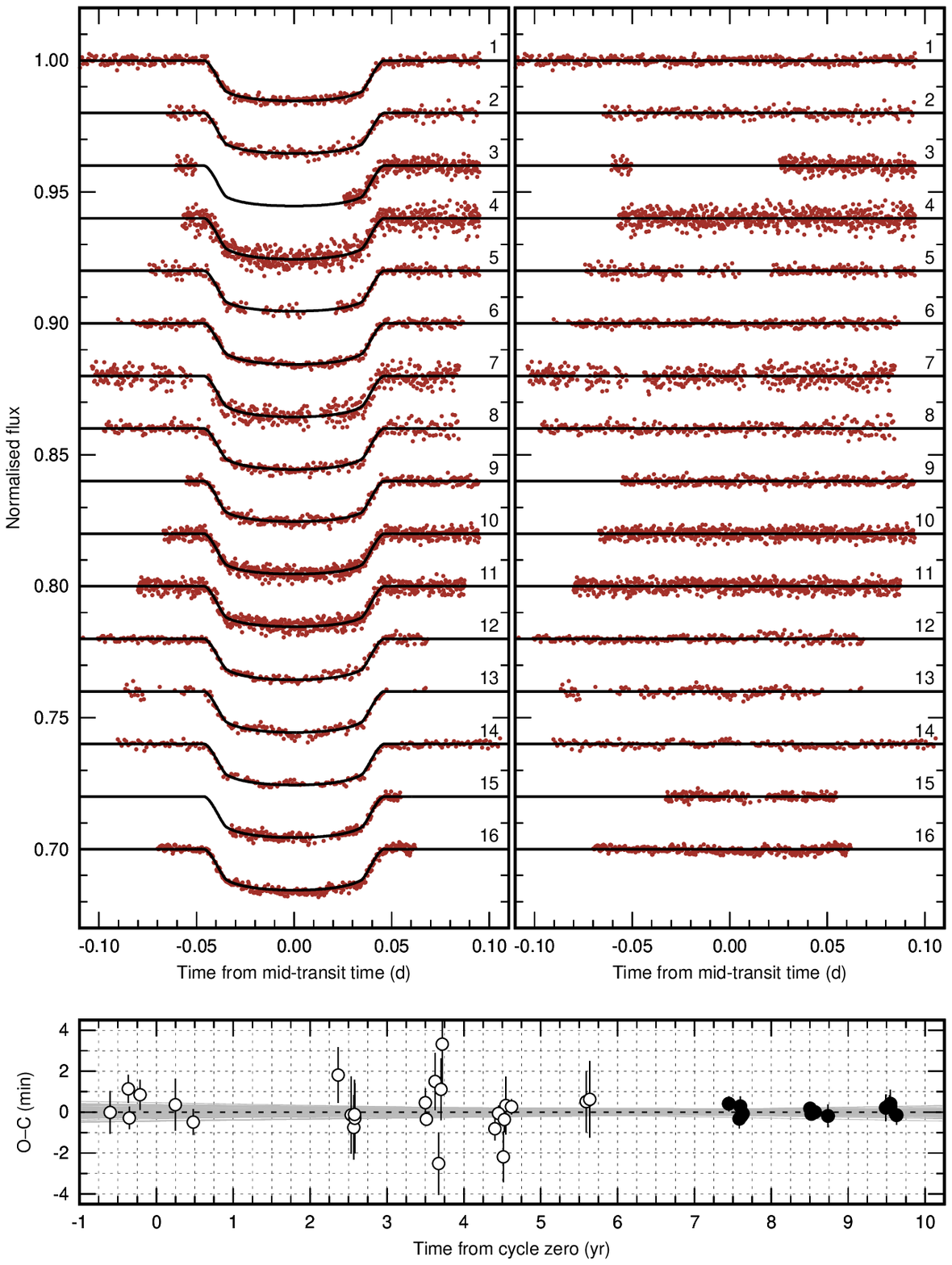}
\end{center}
\FigCap{Upper panels: new transit light curves acquired for HAT-P-23~b with the best-fit models (left) and the residuals (right). Numbers identify the data sets listed in Table~1. Bottom panel: timing residuals against the linear ephemeris for HAT-P-23~b. Values from the new observations are marked with dots, while the redetermined literature values are plotted with open circles. A dashed line marks the zero value. The ephemeris uncertainties are illustrated by grey lines that are drawn for 100 sets of parameters, randomly chosen from the Markov chains.}
\end{figure}

The new and redetermined literature mid-transit times result in the linear ephemeris with the reduced chi-square ($\chi^2_{\rm{red}}$) of 0.7. The timing residuals are plotted in the bottom panel in Fig.~2. Despite the time span of 10 years, no departure from the linear ephemeris can be detected. The trial fit of the quadratic ephemeris results in $\frac{d P_{\rm{orb}}}{d E} = (2.0\pm1.7) \times 10^{-10}$ days per epoch$^{2}$. Thus, the values of $Q'_* < 5.6 \times 10^5$ can be rejected at the 95\% confidence level.

\subsection{KELT-1 b}

Our determinations of the system parameters are based on 9 light curves of millimag and sub-millimag precision (Fig.~3). They are consistent within error bars with the values reported by Siverd \etal (2012) and Baluev \etal (2015). Since the transits are relatively shallow ($\sim$6 ppth), additional photometric time series of sub-millimagnitude precision are desired to place tighter constraints on the system's parameters. We note however that the period uncertainty was reduced thanks to the longer time span of observation.

% FIGURE KELT-1
\begin{figure}[thb]
\begin{center}
\includegraphics[width=1.0\textwidth]{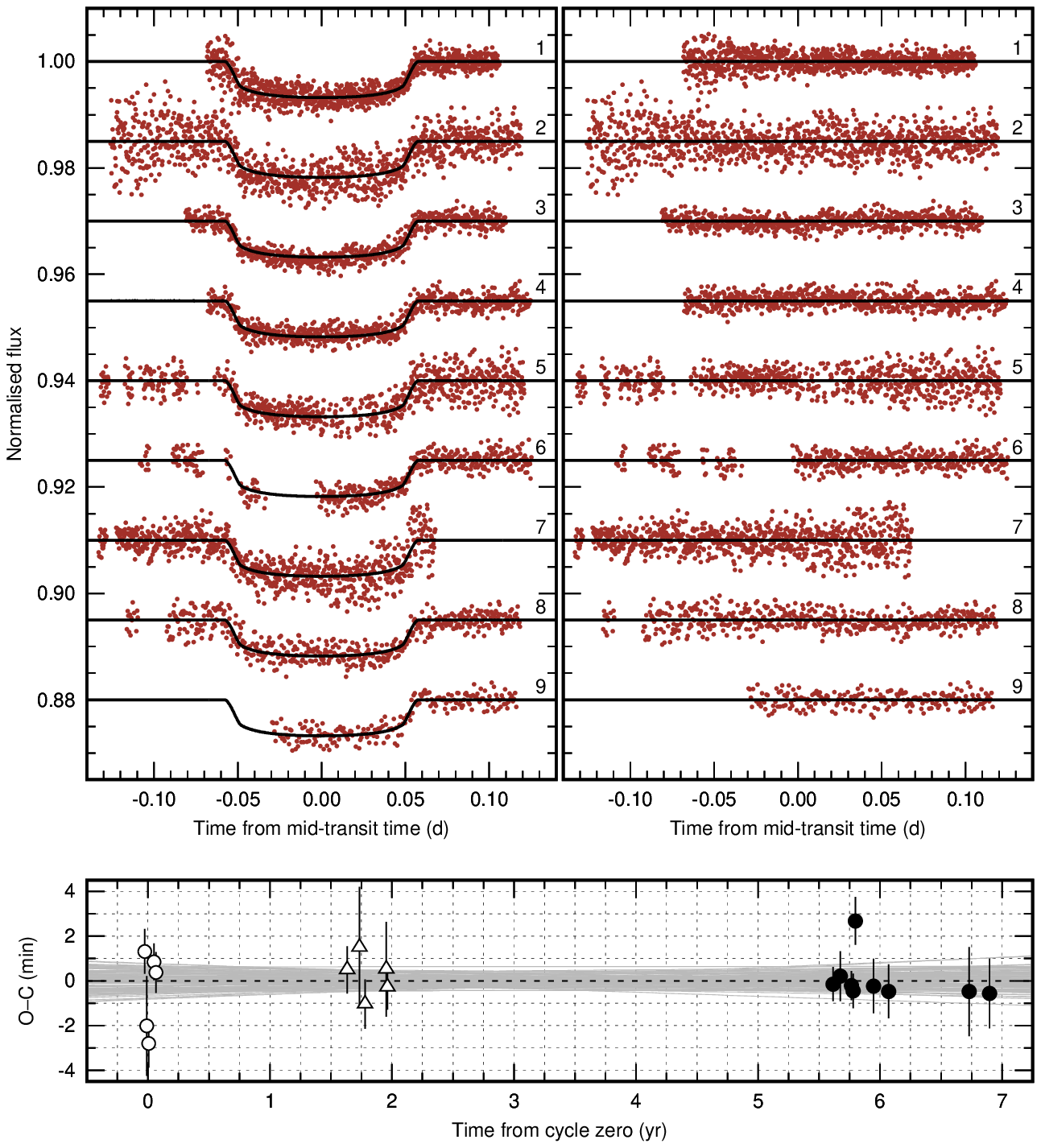}
\end{center}
\FigCap{The same as Fig.~2 but for KELT-1~b. Open circles mark the redetermined mid-transit times only from Siverd \etal (2012), and the supporting amateur data from Baluev \etal (2015), after removing 3-$sigma$ outliers, are plotted with open triangles.}
\end{figure}

In transit timing analysis, the set of mid-transit times from Siverd \etal (2012) and those reported in this paper was enhanced by carefully selected amateur data which were analysed by Baluev \etal (2015). After removing 3-$sigma$ outliers, which degraded the quality of the fit, we finally used 5 mid-transit times with uncertainties taken from the original paper. The model with the linear ephemeris gives $\chi^2_{\rm{red}}$ of 1.1. The trial quadratic model yields $\frac{d P_{\rm{orb}}}{d E} = (-0.1\pm1.3) \times 10^{-9}$ days per epoch$^{2}$. Although further observations will definitely place tighter constraints, current calculations indicate that KELT-1's $Q'_*$ must be greater than $8.4 \times 10^5$ at 95\% confidence.

\subsection{KELT-16 b}

The analysis of 11 new light curves, which are plotted in Fig.~4, allowed us to confirm the values of the transit parameters reported by Oberst \etal (2017). Our observations bring noticeable improvement in the parameters' uncertainties. The orbital period is refined with the accuracy one order of magnitude better than in the discovery paper thanks to the time span of the follow-up observations broadened to 3 years. The linear ephemeris was found to reproduce transit times with $\chi^2_{\rm{red}}$ of 0.7. A trial fit of the quadratic ephemeris results in $\frac{d P_{\rm{orb}}}{d E} = (-0.1\pm1.4) \times 10^{-9}$ days per epoch$^{2}$ that allows us to eliminate $Q'_* < 1.1 \times 10^5$ at the 95\% confidence level. We note however that this weak constraint will be tightened with the increase in the time span of further timing observations.

% FIGURE KELT-16
\begin{figure}[thb]
\begin{center}
\includegraphics[width=1.0\textwidth]{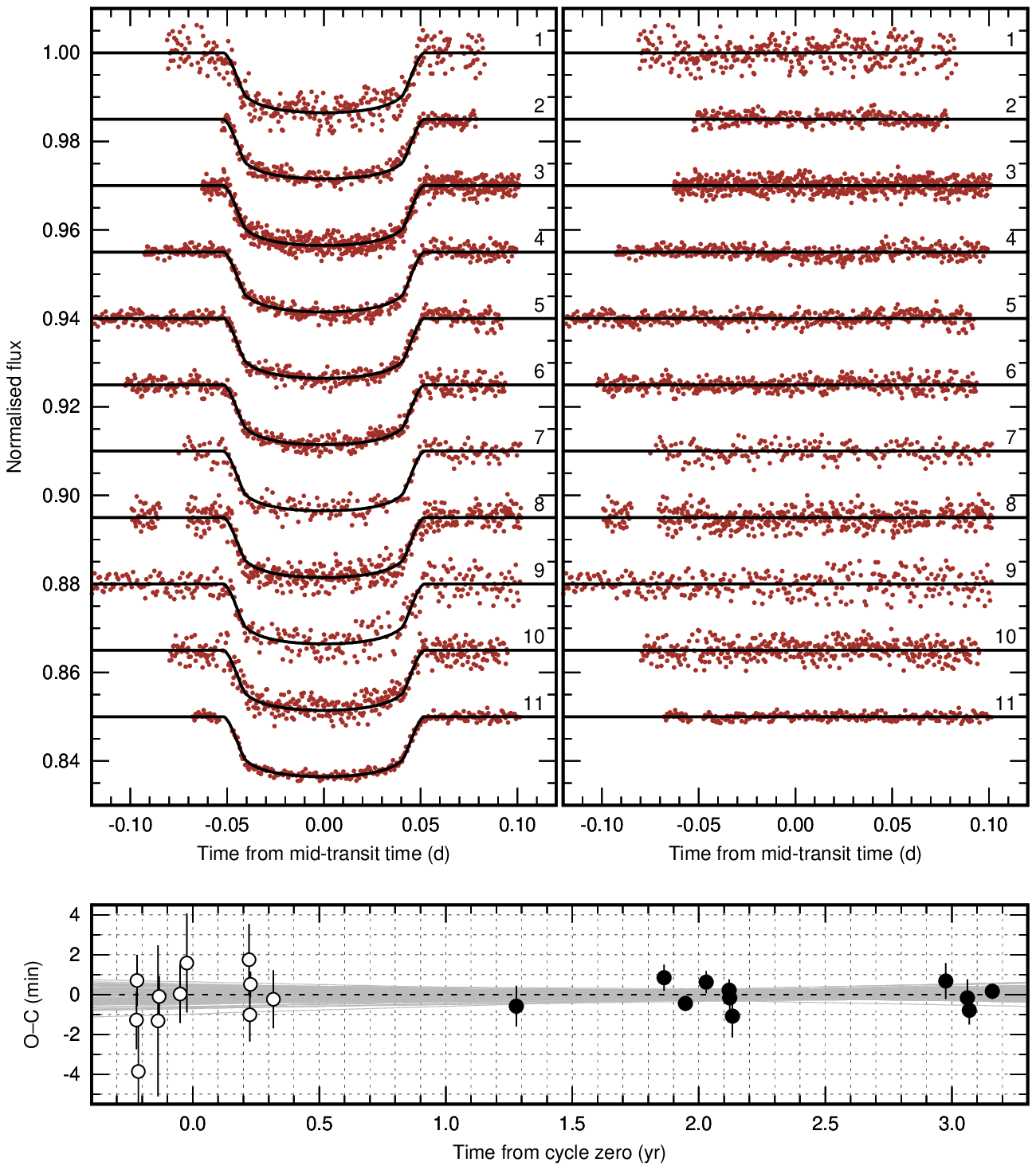}
\end{center}
\FigCap{The same as Fig.~2 but for KELT-16~b.}
\end{figure}

\subsection{WASP-12 b}

Most of the new transit light curves, which are reported for WASP-12~b in this paper {and shown in Fig.~5}, are of excellent quality with the photometric precision below 1 ppth per minute. This allowed us to refine transit parameters with uncertainties comparable to or even better than those reported in Maciejewski \etal (2013). The values of these parameters were found to be consistent with the literature ones well within the 1$\sigma$ level.   

To keep the homogeneity of the timing analysis, we used the literature data reanalysed in Maciejewski \etal (2016). As it is shown in the bottom panel in Fig.~5, the new transit times follow the quadratic ephemeris very well. The best-fit model has $\chi^2_{\rm{red}}$ of 0.9 and yields 
\begin{equation}
\frac{d P_{\rm{orb}}}{d E} = (-9.67\pm0.73) \times 10^{-10}\quad \rm{days~per~epoch}^{2} \, . \;
\end{equation} 
This value is consistent within error bars with and more precise than values of $(-8.9\pm1.4) \times 10^{-10}$ and $(-10.2\pm1.1) \times 10^{-10}$ days per epoch$^{2}$ reported by Maciejewski \etal (2016) and Patra \etal (2017), respectively. Following eq.~(5), we obtained
\begin{equation}
Q'_* = (1.82 \pm 0.32) \times 10^{5} \, , \;
\end{equation} 
where the value of $M_p/M_{*}$ was taken from Bonomo \etal (2017) and the remaining parameters from this study (Table~2).

% FIGURE WASP-12
\begin{figure}[thb]
\begin{center}
\includegraphics[width=1.0\textwidth]{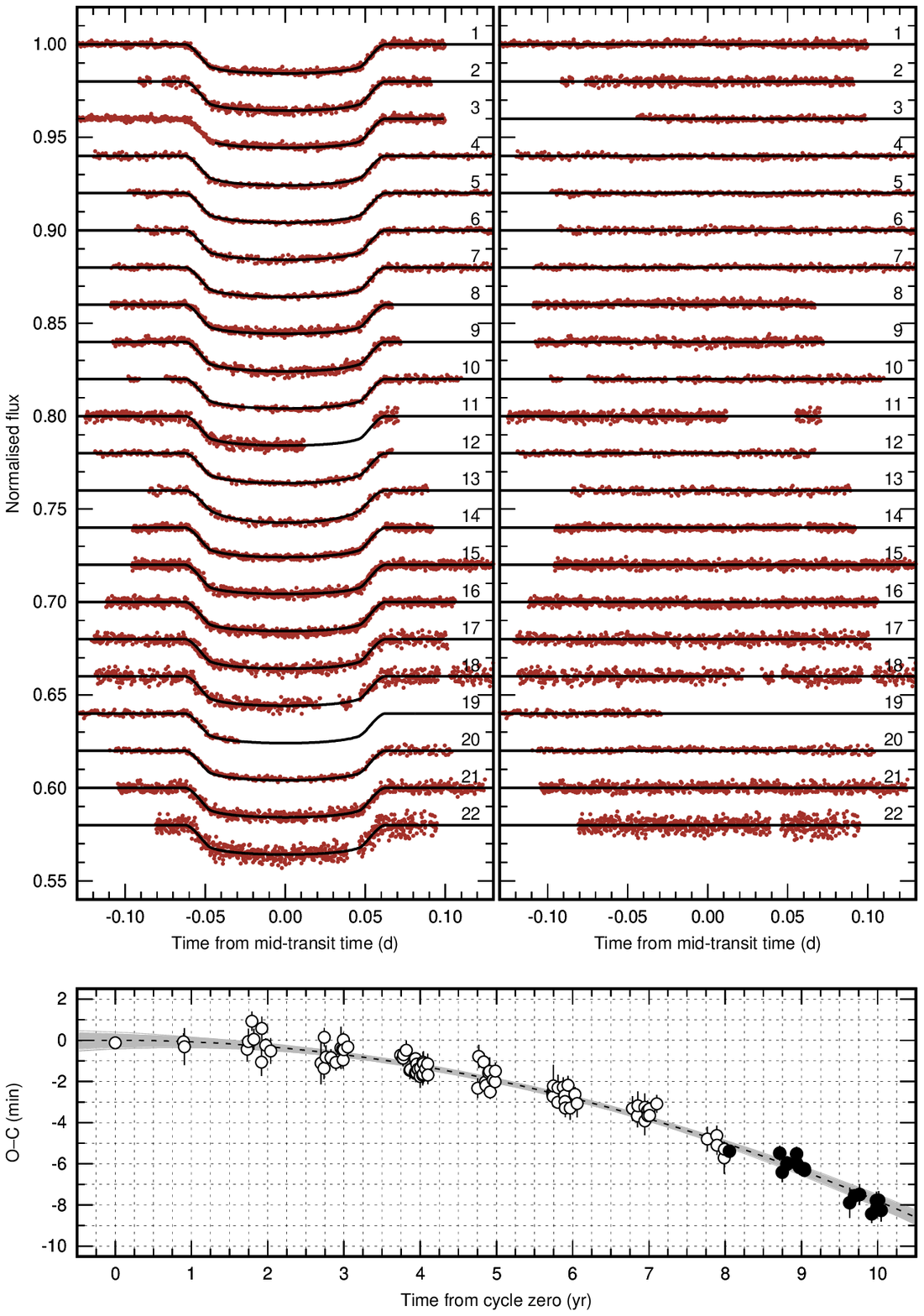}
\end{center}
\FigCap{The same as Fig.~2 but for WASP-12~b with the difference that the dashed line in the bottom panel displays the best-fit quadratic trend in transit times, and the grey lines sketch the uncertainties of the quadratic ephemeris.}
\end{figure}

\subsection{WASP-33 b}

% FIGURE WASP-33
\begin{figure}[thb]
\begin{center}
\includegraphics[width=1.0\textwidth]{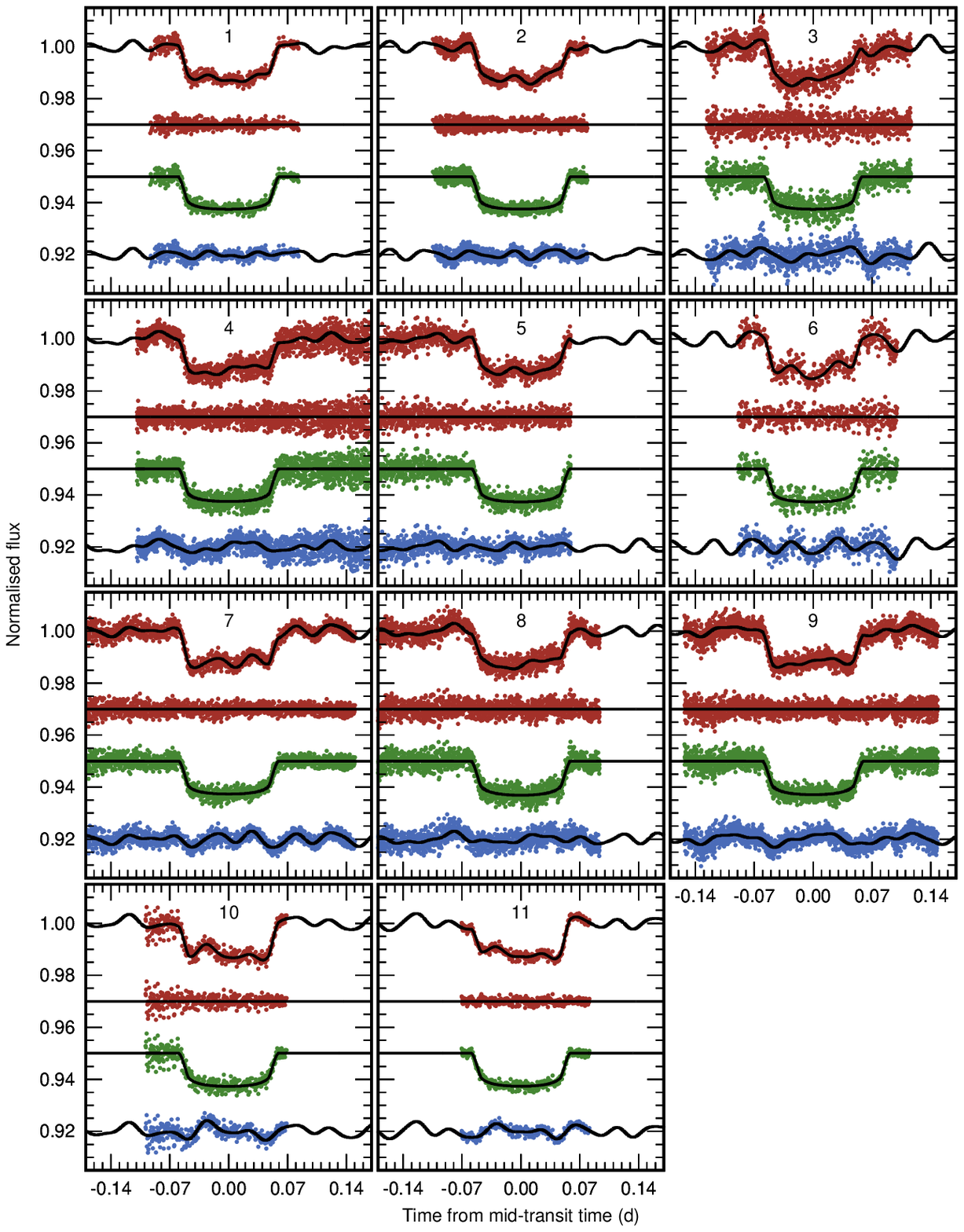}
\end{center}
\FigCap{Transit light curves acquired for WASP-33~b. In each panel, there are plotted (from top to bottom): raw light curves, the final residuals, and flux changes induced by planetary transits and stellar variation. Dots show individual measurements while the best-fit models are sketched with continuous lines. Numbers identify the data sets listed in Table~1.} 
\end{figure}

\begin{figure}[thb]
\begin{center}
\includegraphics[width=1.0\textwidth]{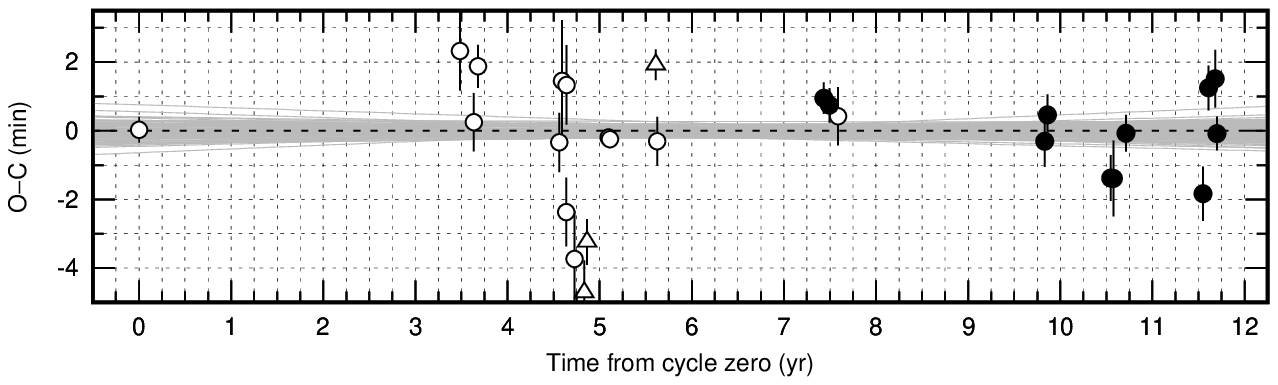}
\end{center}
\FigCap{Timing residuals against the linear ephemeris  for WASP-33~b. Values from the new observations are marked with dots, while the literature values, which were used in this study, are plotted with open circles. The 3-$\sigma$ outliers, which were rejected in the analysis, are shown with open triangles. The dashed line marks zero value. The ephemeris uncertainties are illustrated by grey lines that are drawn for 100 sets of parameters, randomly chosen from the Markov chains.} 
\end{figure}

Although the complex photometric variability of WASP-33 remains outside the scope of this paper, it affects the transit light curves and must be included in the modelling to determine correct values of the transit parameters (e.g.\ von Essen \etal 2014). Since variable flux baseline is not implemented in TAP, additional steps were introduced to our data analysis procedure. First, a trial transit model with the parameters taken from von Essen \etal (2014) was fitted to a raw light curve. Then, the stellar variation was modelled using the data outside the transit and at the flat bottom phase. The ingress and egress were masked out with the margins of $\pm5$ minutes to eliminate those portions of the data that could be affected by the transit morphology. In the next step, the residuals were modelled with the function 
\begin{equation}
f(t) = a_0 + a_1t + a_2 t^2 + \sum_{i=1}^{8} A_i \sin \left(\nu_i t +\phi_i \right) \, , \;
\end{equation} 
where $a_0$, $a_1$, and $a_2$ are the coefficients of a parabolic trend against time $t$, and $A_i$, $\nu_i$, and $\phi_i$ are the amplitudes, frequencies, and phases of sinusoids which correspond to 8 pulsation frequencies reported by von Essen \etal (2014). The parameters $A_i$ and $\nu_i$ were fixed at the literature values because our photometric time series were not long enough to redetermined them in a reliable way. Since the phase of pulsations can change, we allowed $\phi_i$ to float. Because of the complex variation signal, the raw light curves were subjected to de-trending against time only. The parabolic component was used to account for any possible trends caused by changes in airmass or weather conditions, or induced by instrumental effects. The best-fit of $f(t)$ was found with the least squares method. The model of stellar variations was subtracted in magnitudes from the raw light curve which was then normalised to out-of-transit flux. The standard transit modelling procedure was performed to determine system parameters. The individual light curves with the stellar variation and the transit component split are shown in Fig.~6.

Our determinations of the transit parameters were compared to those literature studies which took stellar variability into account. As it can be seen in Table~2, our results agree within 1-$\sigma$ with most of the values reported by Kov\'acs \etal (2013), von Essen \etal (2014), and Zhang \etal (2018). We note that the value of $R_p/R_*$ obtained by Kov\'acs \etal (2013) appears to be  significantly overestimated as compared to the other determinations.

In the transit timing analysis, we used mid-transit times from Collier Cameron \etal (2010), von Essen \etal (2014), Johnson \etal (2015), and Zhang \etal (2018) which are based on professional photometric observations. As some fraction of the data may still be affected by stellar variability, a 3-$\sigma$ clipping was iteratively applied while refining the transit ephemeris. The final set of mid-transit times used is listed in Table~7, and the timing residuals against the linear ephemeris are plotted in Fig.~7. Our new observations extend the time span of WASP-33~b's monitoring and hence allow us to refine the planet's transit ephemeris with unprecedented accuracy. The goodness of the fit $\chi^2_{\rm{red}}=2.5$ was found to be degraded by determinations affected by the stellar variability. The trial quadratic ephemeris gives $\frac{d P_{\rm{orb}}}{d E} = (-0.3\pm1.3) \times 10^{-10}$ days per epoch$^{2}$, hence we conclude that $Q'_*$ of WASP-33 must be greater than $8.8 \times 10^5$ with 95\% confidence.

\subsection{WASP-103 b}

Our analysis of the new transit light curves, which are displayed in Fig.~8, resulted in the best-fit model that reproduces the literature values of transit parameters well within 1--2$\sigma$. For timing purposes, we reanalysed the transit photometry from Gillon \etal (2014), Southworth \etal (2015), Delrez \etal (2018), Turner \etal (2017), and Lendl \etal (2017). Our new timing dataset almost doubles the time span covered with observations. Thanks to this, the period uncertainty was reduced by one order of magnitude. The model assuming pure Keplerian motion of the planet was found to reproduce the timing observations with $\chi^2_{\rm{red}}$ of 1.1. Applying the quadratic ephemeris gives $\frac{d P_{\rm{orb}}}{d E} = (1.03\pm0.36) \times 10^{-9}$ days per epoch$^{2}$. This positive value of $\frac{d P_{\rm{orb}}}{d E}$, which is consistent with zero within the $\sim$3$\sigma$ level, prevents us from placing a constraint on $Q'_*$ at the 95\% confidence level. We find, however, that $Q'_*$ must be greater than $10^6$ at the 99.96\% (i.e., $3.5 \sigma$). While looking at the bottom panel in Fig.~8, one can note that four transits observed $\sim$2 years after cycle zero appeared $\sim$1 minute earlier than the linear ephemeris predicted. Although those mid-transit times were re-determined using datasets from different sources (Delrez \etal 2017, Turney \etal 2017, Lendl \etal 2017), their deviations are caused rather by accidental systematic errors than by a real astrophysical signal. No other deviations from the linear ephemeris can be detected in following observations.

\begin{figure}[thb]
\begin{center}
\includegraphics[width=1.0\textwidth]{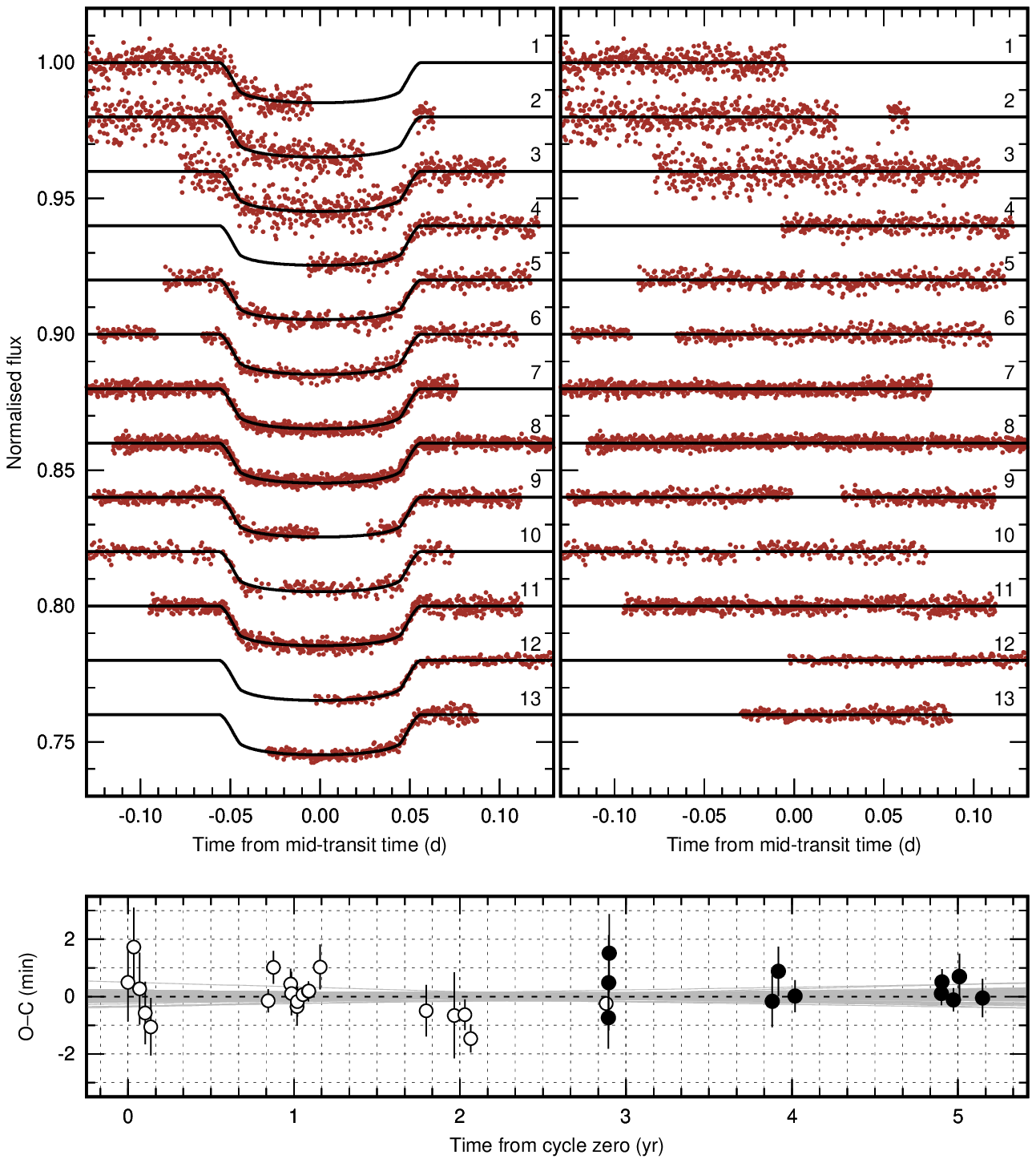}
\end{center}
\FigCap{The same as Fig.~2 but for WASP-103~b.}
\end{figure}

\section{Concluding discussion}

Employing the formalism of Ivanov \etal (2013), Chernov \etal (2017) demonstrated that the rate of the orbital shrinkage of WASP-12~b might be consistent with theoretical predictions which assume the host star is a dwarf. In this scenario, the WASP-12 system would be nowadays observed in the final stage of its existence, which appears to be unlikely (Patra \etal 2017). On the other hand, Weinberg \etal (2017) found that the observed rate of orbital decay could only be explained if WASP-12 were a subgiant -- a star during the transition phase between the end of the main-sequence stage and the beginning of stable hydrogen burning in a shell on the red-giant branch. This scenario is supported by the stellar properties which are consistent with a $\sim$1.2 $M_{\odot}$ subgiant star. Models of the internal structure of subgiants predict that the efficiency of the tidal dissipation is boosted by several orders of magnitude due to nonlinear wave-breaking of the dynamical tide near the star's centre (Barker \& Ogilvie 2010). If this mechanism operates in WASP-12, the calculations of Barker \& Ogilvie (2010) yield $Q'_* \approx 1.9 \times 10^5$ that is in an excellent agreement with our empirical $Q'_* \approx 1.8 \times 10^5$. The high value of $Q'_* \gtrsim 10^8$ (Barker \& Ogilvie 2009) would have prevented the planet from spiralling inward over the course of $\sim$4 Gyr of WASP-12's evolution on the main sequence. The evolutionary changes in the star's interior structure would then trigger a rapid orbital decay that is observed nowadays (Barker \& Ogilvie 2010, Weinberg \etal 2017). Our new transit timing observations fully support this scenario. We note, however, that further observations, including precise occultation timing, are still required to verify this interpretation.

HAT-P-23 has an effective temperature of $\sim$5900 K (Bakos \etal 2011) and is the most Sun-like star in our sample. Tidal dissipation occurs mainly in the convective layer. However, the weakly non-linear wave interactions close to the centre of a solar-type star may still play an important role, enhancing the rate of tidal dissipation (Ogilvie \& Lin 2007, Essick \& Weinberg 2016). For the HAT-P-23 system, the model of Essick \& Weinberg (2016) predicts $Q'_* \approx 6.7 \times 10^5$ that is not rejected by our empirical constraint of $Q'_* \gtrsim 5.6 \times 10^5$. On the other hand, Penev \etal (2018) obtained $Q'_* \approx 3 \times 10^6$ assuming that HAT-P-23 with its rotational speed observed nowadays has been spin up by tides being raised by its hot giant planet. Further precise transit timing data are expected to empirically verify the proposed models. 

Our calculations show that the massive KELT-1~b is the best candidate for the detection of planet-star tidal interactions. Our preliminary constraint allows us to reject $Q'_* \lesssim 8.4 \times 10^5$ that is consistent with theoretical predictions for F-type dwarfs (Barker \& Ogilvie 2009), as well as with  $Q'_* \geq 1 \times 10^6$ obtained for the similar star WASP-18 (Wilkins \etal 2017). On the other hand, estimates of the stellar rotation period suggest that the star and its companion might be in a state of tidal equilibrium (Siverd \etal 2012). If both the rotation period of the host star and the orbital period of KELT-1~b are synchronised there is no tidal lag and a transfer of angular momentum due to tidal interactions does not take place. Further transit timing observations may provide tighter constraints on $Q'_*$ in order to address the questions on the dynamical state of this system. 

KELT-16 and WASP-103, just like WASP-12, have effective temperatures close to the Kraft break which separates stars with radiative and convective envelopes (Kraft 1967). Assuming that both stars are dwarfs, they are expected to posses convective cores and thin -- if any -- convective outer layers. For such interior structure, the tidal dissipation is expected to be rather weak (Penev \etal 2018). For KELT-16, the time span of timing monitoring is just 3 years, and is definitely too short to place a reasonable constraint on $Q'_*$. The case of WASP-103 shows however that the expected value of $Q'_*$ is likely $\gtrsim 10^6$. This finding is strengthened by spectroscopic observations showing that both hosts, unlike WASP-12, are main sequence stars (Oberst \etal 2017, Gillon \etal 2014). Further transit timing observations shall provide tighter constraints on $Q'_*$ for those stars.

With an effective temperature of $\sim$7300 K (Collier Cameron \etal 2010), WASP-33 has a convective core and a radiative envelope. Observations of binary stars (Khaliullin \& Khaliullina 2010) confirm that dynamical tides, which are excited at the boundary between the convective core and radiative envelope, may be dissipated through the radiative damping of internal gravity waves in the non-adiabatic layers located near the stellar surface (e.g. Zahn 1975). The efficiency of this dissipation mechanism may be however orders of magnitude lower than for Sun-like stars with extensive convective envelopes (Zahn 1977). The WASP-33 system is a great laboratory for testing the rate of tidal dissipation in A-type stars. Although the photometric variability of the host star makes accurate transit timing challenging, our observations demonstrate that it is possible to determine mid-transit times with sub-minute precision.

%%%%%%%%%%%%%%%%%%%%%%%%%%%%%%%%%%%%%%%%%%%%%%%%%%%%%%%%%%%%%%%%%%%%%%

\Acknow{We thank the referee for valuable comments which improved the paper. We also thank Dr. Laetitia Delrez and Dr. Pedro Sada for sharing the WASP-103 and HAT-P-23 light curves with us. GM and MS acknowledge the financial support from the National Science Centre, Poland through grant no. 2016/23/B/ST9/00579. MF acknowledges financial support from grants AYA2014-54348-C3-1-R and AYA2016-79425-C3-3-P of the Spanish Ministry of Economy and Competitiveness (MINECO), co-funded with EU FEDER funds. DD acknowledges the financial support of projects DN 08-1/2016, and DN 08-20/2016 of National Science Foundation of Bulgarian Ministry of education and science as well as by project RD 08-142 of Shumen University. CvE acknowledges funding for the Stellar Astrophysics Centre, which is provided by The Danish National Research Foundation (Grant agreement no.: DNRF106). DM acknowledges support from the National Science Centre (NCN) grant no. 2016/21/B/ST9/01126. A part of this paper is the result of the exchange and joint research project {\em Spectral and photometric studies of variable stars} between the Polish and Bulgarian Academies of Sciences. This project has received funding from the European Union's Horizon 2020 research and innovation programme under grant agreement No 730890. This material reflects only the authors views and the Commission is not liable for any use that may be made of the information contained therein. This research is based on (1) data obtained at the 1.5m telescope of the Sierra Nevada Observatory (Spain), which is operated by the Consejo Superior de Investigaciones Cient\'{\i}ficas (CSIC) through the Instituto de Astrof\'{\i}sica de Andaluc\'{\i}a, (2) data collected with telescopes at the Rozhen National Astronomical Observatory, (3) observations made with the Liverpool Telescope operated on the island of La Palma by Liverpool John Moores University in the Spanish Observatorio del Roque de los Muchachos of the Instituto de Astrof\'isica de Canarias with financial support from the UK Science and Technology Facilities Council, (4) observations made with the Gran Telescopio Canarias (GTC), installed in the Spanish Observatorio del Roque de los Muchachos of the Instituto de Astrof\'isica de Canarias, in the island of La Palma, and (5) observations obtained with telescopes of the University Observatory Jena, which is operated by the Astrophysical Institute of the Friedrich-Schiller-University.}

\newpage

\vglue-3mm
\centerline{\bf Appendix}
\vskip1mm
\centerline{\bf Mid-transit times used for timing studies.} 
\centerline{\small Tables 3--8 in a machine-readable format are available at}
\centerline{\it http://www.home.umk.pl/\~{}gmac/TTV/doku.php?id=download}
\centerline{\small or via CDS.}

\MakeTable{ r l l l c l}{12.5cm}{New and redetermined literature mid-transit times for HAT-P-23 b.}
{\hline
 Epoch$^{\rm{a}}$ & $T_{\rm{mid}}$ (BJD$_{\rm{TDB}}$) & \multicolumn{1}{c}{$+\sigma$ (d)} & \multicolumn{1}{c}{$-\sigma$ (d)} & $N_{\rm{tr}}$ & Light curve source\\
\hline
-181 & 2454632.73289 & 0.00073 & 0.00072 & 1 & Bakos \etal (2011) \\
-110 & 2454718.84863 & 0.00047 & 0.00048 & 1 & Bakos \etal (2011) \\
-106 & 2454723.69918 & 0.00039 & 0.00039 & 1 & Bakos \etal (2011) \\
-64 & 2454774.64120 & 0.00056 & 0.00051 & 1 & Bakos \etal (2011) \\
74 & 2454942.01919 & 0.00082 & 0.00088 & 1 & Bakos \etal (2011) \\
144 & 2455026.92065 & 0.00044 & 0.00043 & 1 & Bakos \etal (2011) \\
712 & 2455715.84171 & 0.00101 & 0.00095 & 1 & Ram\'on-Fox \& Sada (2013) \\
763 & 2455777.6976 & 0.0013 & 0.0014 & 1 & Ram\'on-Fox \& Sada (2013) \\
773 & 2455789.8260 & 0.0011 & 0.0011 & 1 & Ram\'on-Fox \& Sada (2013) \\
776 & 2455793.4651 & 0.0012 & 0.0011 & 1 & Ciceri \etal (2015) \\
777 & 2455794.6779 & 0.0012 & 0.0012 & 1 & Ram\'on-Fox \& Sada (2013) \\
1054 & 2456130.64792 & 0.00051 & 0.00050 & 1 & Ciceri \etal (2015) \\
1058 & 2456135.49891 & 0.00022 & 0.00023 & 1 & Ciceri \etal (2015) \\
1092 & 2456176.73833 & 0.00102 & 0.00098 & 1 & Sada \& Ram\'on-Fox (2016) \\
1106 & 2456193.7160 & 0.0011 & 0.0011 & 1 & Sada \& Ram\'on-Fox (2016) \\
1115 & 2456204.6345 & 0.0011 & 0.0011 & 1 & Sada \& Ram\'on-Fox (2016) \\
1120 & 2456210.7004 & 0.0013 & 0.0015 & 1 & Sada \& Ram\'on-Fox (2016) \\
1326 & 2456460.55214 & 0.00040 & 0.00039 & 1 & Ciceri \etal (2015) \\
1340 & 2456477.53307 & 0.00027 & 0.00027 & 1 & Ciceri \etal (2015) \\
1359 & 2456500.57644 & 0.00086 & 0.00090 & 1 & Ciceri \etal (2015) \\
1363 & 2456505.42925 & 0.00044 & 0.00042 & 1 & Ciceri \etal (2015) \\
1369 & 2456512.70704 & 0.00103 & 0.00099 & 1 & Sada \& Ram\'on-Fox (2016) \\
1391 & 2456539.39051 & 0.00025 & 0.00026 & 3$^{\rm{b}}$ & Ciceri \etal (2015) \\
1684 & 2456894.7664 & 0.0010 & 0.0010 & 1 & Sada \& Ram\'on-Fox (2016) \\
1698 & 2456911.7469 & 0.0013 & 0.0013 & 1 & Sada \& Ram\'on-Fox (2016) \\
2242 & 2457571.55693 & 0.00025 & 0.00026 & 1 & this paper \\
2284 & 2457622.49765 & 0.00033 & 0.00034 & 2 & this paper \\
2288 & 2457627.34962 & 0.00034 & 0.00033 & 1 & this paper \\
2293 & 2457633.41373 & 0.00016 & 0.00017 & 3 & this paper \\
2298 & 2457639.47823 & 0.00029 & 0.00029 & 1 & this paper \\
2561 & 2457958.46753 & 0.00021 & 0.00021 & 1 & this paper \\
2566 & 2457964.53178 & 0.00021 & 0.00021 & 1 & this paper \\
2580 & 2457981.51225 & 0.00024 & 0.00024 & 1 & this paper \\
2631 & 2458043.36933 & 0.00039 & 0.00039 & 1 & this paper \\
2857 & 2458317.48193 & 0.00047 & 0.00046 & 1 & this paper \\
2871 & 2458334.46241 & 0.00034 & 0.00035 & 1 & this paper \\
2875 & 2458339.31403 & 0.00045 & 0.00047 & 1 & this paper \\
2899 & 2458368.42291 & 0.00033 & 0.00033 & 1 & this paper \\
\hline
\multicolumn{6}{l}{$N_{\rm{tr}}$ is the number of individual light curves taken to determine $T_{\rm{mid}}$.} \\
\multicolumn{6}{l}{$^{\rm{a}}$ Epoch 0 set for $T_0$ from Bakos \etal (2011).}\\
\multicolumn{6}{l}{$^{\rm{b}}$ Light curve acquired in the Thuan-Gunn $u$ filter was skipped due to low quality} \\
\multicolumn{6}{l}{that degraded the fit.} \\
}

\MakeTable{ r l l l c l}{12.5cm}{New and redetermined literature mid-transit times for KELT-1 b.}
{\hline
 Epoch$^{\rm{a}}$ & $T_{\rm{mid}}$ (BJD$_{\rm{TDB}}$) & \multicolumn{1}{c}{$+\sigma$ (d)} & \multicolumn{1}{c}{$-\sigma$ (d)} & $N_{\rm{tr}}$ & Light curve source\\
\hline
-8 & 2455899.55385 & 0.00071 & 0.00070 & 2 & Siverd \etal (2012) \\
-3 & 2455905.6390 & 0.0016 & 0.0015 & 1 & Siverd \etal (2012) \\
2 & 2455911.72593 & 0.00075 & 0.00072 & 1 & Siverd \etal (2012) \\
15 & 2455927.55589 & 0.00057 & 0.00057 & 1 & Siverd \etal (2012) \\
20 & 2455933.64303 & 0.00064 & 0.00063 & 1 & Siverd \etal (2012) \\
490 & 2456505.86528$^{\rm{b}}$ & 0.00073$^{\rm{b}}$ & 0.00073$^{\rm{b}}$ & 1 & Baluev \etal (2015) \\
520 & 2456542.3908$^{\rm{b}}$ & 0.0019$^{\rm{b}}$ & 0.0019$^{\rm{b}}$ & 1 & Baluev \etal (2015) \\
534 & 2456559.43395$^{\rm{b}}$ & 0.00077$^{\rm{b}}$ & 0.00077$^{\rm{b}}$ & 1 & Baluev \etal (2015) \\
586 & 2456622.7447$^{\rm{b}}$ & 0.0015$^{\rm{b}}$ & 0.0015$^{\rm{b}}$ & 1 & Baluev \etal (2015) \\
589 & 2456626.39665$^{\rm{b}}$ & 0.00071$^{\rm{b}}$ & 0.00071$^{\rm{b}}$ & 1 & Baluev \etal (2015) \\
1684 & 2457959.55263 & 0.00054 & 0.00053 & 1 & this paper \\
1702 & 2457981.46777 & 0.00078 & 0.00077 & 1 & this paper \\
1730 & 2458015.55731 & 0.00046 & 0.00046 & 1 & this paper \\
1734 & 2458020.42711 & 0.00054 & 0.00055 & 1 & this paper \\
1739 & 2458026.51676 & 0.00073 & 0.00074 & 1 & this paper \\
1784 & 2458081.30196 & 0.00085 & 0.00084 & 1 & this paper \\
1821 & 2458126.34908 & 0.00084 & 0.00085 & 1 & this paper \\
2019 & 2458367.4129 & 0.0014 & 0.0013 & 1 & this paper \\
2069 & 2458428.2875 & 0.0011 & 0.0011 & 1 & this paper \\
\hline
\multicolumn{6}{l}{$N_{\rm{tr}}$ is the number of individual light curves taken to determine $T_{\rm{mid}}$.} \\
\multicolumn{6}{l}{$^{\rm{a}}$ Epoch 0 set for $T_0$ from Siverd \etal (2012).}\\
\multicolumn{6}{l}{$^{\rm{b}}$ Values taken from the source paper.}\\
}

\MakeTable{ r l l l c l}{12.5cm}{New and redetermined literature mid-transit times for KELT-16 b.}
{\hline
 Epoch$^{\rm{a}}$ & $T_{\rm{mid}}$ (BJD$_{\rm{TDB}}$) & \multicolumn{1}{c}{$+\sigma$ (d)} & \multicolumn{1}{c}{$-\sigma$ (d)} & $N_{\rm{tr}}$ & Light curve source\\
\hline
-84 & 2457165.85142 & 0.00101 & 0.00097 & 1 & Oberst \etal (2017) \\
-83 & 2457166.82179 & 0.00084 & 0.00089 & 2 & Oberst \etal (2017) \\
-81 & 2457168.7566 & 0.0017 & 0.0020 & 1 & Oberst \etal (2017) \\
-52 & 2457196.8592 & 0.0026 & 0.0029 & 1 & Oberst \etal (2017) \\
-50 & 2457198.79802 & 0.00070 & 0.00074 & 4 & Oberst \etal (2017) \\
-19 & 2457228.8369 & 0.0010 & 0.0010 & 1 & Oberst \etal (2017) \\
-9 & 2457238.5279 & 0.0018 & 0.0017 & 1 & Oberst \etal (2017) \\
84 & 2457328.6444 & 0.0014 & 0.0012 & 1 & Oberst \etal (2017) \\
85 & 2457329.61146 & 0.00094 & 0.00091 & 1 & Oberst \etal (2017) \\
86 & 2457330.58151 & 0.00046 & 0.00045 & 4 & Oberst \etal (2017) \\
120 & 2457363.52676 & 0.00101 & 0.00097 & 2 & Oberst \etal (2017) \\
482 & 2457714.30206 & 0.00071 & 0.00070 & 1 & this paper \\
702 & 2457927.48156 & 0.00048 & 0.00046 & 1 & this paper \\
734 & 2457958.48844 & 0.00026 & 0.00026 & 1 & this paper \\
765 & 2457988.52797 & 0.00038 & 0.00039 & 1 & this paper \\
799 & 2458021.47346 & 0.00039 & 0.00038 & 1 & this paper \\
800 & 2458022.44219 & 0.00047 & 0.00045 & 1 & this paper \\
804 & 2458026.31752 & 0.00074 & 0.00068 & 1 & this paper \\
1122 & 2458334.45858 & 0.00059 & 0.00063 & 1 & this paper \\
1154 & 2458365.46578 & 0.00064 & 0.00062 & 1 & this paper \\
1157 & 2458368.37232 & 0.00048 & 0.00047 & 1 & this paper \\
1191 & 2458401.31876 & 0.00028 & 0.00027 & 1 & this paper \\
\hline
\multicolumn{6}{l}{$N_{\rm{tr}}$ is the number of individual light curves taken to determine $T_{\rm{mid}}$.} \\
\multicolumn{6}{l}{$^{\rm{a}}$ Epoch 0 set for $T_0$ from Oberst \etal (2017).}\\
}

\MakeTable{ r l l l c}{12.5cm}{New mid-transit times for WASP-12 b.}
{\hline
 Epoch$^{\rm{a}}$ & $T_{\rm{mid}}$ (BJD$_{\rm{TDB}}$) & \multicolumn{1}{c}{$+\sigma$ (d)} & \multicolumn{1}{c}{$-\sigma$ (d)} & $N_{\rm{tr}}$ \\
\hline
2696 & 2457451.44617 & 0.00021 & 0.00021 & 1 \\
2916 & 2457691.55888 & 0.00025 & 0.00025 & 1 \\
2927 & 2457703.56388 & 0.00034 & 0.00032 & 1 \\
2948 & 2457726.48400 & 0.00028 & 0.00028 & 1 \\
2949 & 2457727.57547 & 0.00023 & 0.00023 & 1 \\
2990 & 2457772.32407 & 0.00024 & 0.00024 & 1 \\
2991 & 2457773.41517 & 0.00022 & 0.00021 & 1 \\
3003 & 2457786.51210 & 0.00026 & 0.00026 & 2 \\
3024 & 2457809.43190 & 0.00018 & 0.00018 & 2 \\
3025 & 2457810.52327 & 0.00020 & 0.00021 & 1 \\
3223 & 2458026.62368 & 0.00051 & 0.00056 & 1 \\
3245 & 2458050.63519 & 0.00023 & 0.00023 & 1 \\
3266 & 2458073.55509 & 0.00022 & 0.00022 & 1 \\
3267 & 2458074.64651 & 0.00034 & 0.00034 & 1 \\
3320 & 2458132.49121 & 0.00031 & 0.00029 & 1 \\
3341 & 2458155.41152 & 0.00030 & 0.00031 & 2 \\
3342 & 2458156.50267 & 0.00031 & 0.00032 & 1 \\
3351 & 2458166.32575 & 0.00031 & 0.00034 & 1 \\
3362 & 2458178.33104 & 0.00037 & 0.00038 & 1 \\
\hline
\multicolumn{5}{l}{$N_{\rm{tr}}$ is the number of individual light curves taken to determine $T_{\rm{mid}}$.} \\
\multicolumn{5}{l}{$^{\rm{a}}$ Epoch 0 set for $T_0$ from Hebb \etal (2009).}\\
}

\MakeTable{ r l l l c l}{12.5cm}{New and literature mid-transit times for WASP-33 b.}
{\hline
 Epoch$^{\rm{a}}$ & $T_{\rm{mid}}$ (BJD$_{\rm{TDB}}$) & \multicolumn{1}{c}{$+\sigma$ (d)} & \multicolumn{1}{c}{$-\sigma$ (d)} & $N_{\rm{tr}}$ & Light curve source\\
\hline
   0 & 2454163.22451$^{\rm{b}}$ & 0.00026$^{\rm{b}}$ & 0.00026$^{\rm{b}}$ & 1 & Collier Cameron \etal (2010) \\
1043 & 2455435.55145$^{\rm{b}}$ & 0.00086$^{\rm{b}}$ & 0.00080$^{\rm{b}}$ & 1 & von Essen \etal (2014) \\
1088 & 2455490.44420$^{\rm{b}}$ & 0.00059$^{\rm{b}}$ & 0.00058$^{\rm{b}}$ & 1 & von Essen \etal (2014) \\
1102 & 2455507.52352$^{\rm{b}}$ & 0.00045$^{\rm{b}}$ & 0.00043$^{\rm{b}}$ & 1 & von Essen \etal (2014) \\
1366 & 2455829.56790$^{\rm{b}}$ & 0.00060$^{\rm{b}}$ & 0.00062$^{\rm{b}}$ & 1 & von Essen \etal (2014) \\
1375 & 2455840.5480$^{\rm{b}}$ & 0.0012$^{\rm{b}}$ & 0.0012$^{\rm{b}}$ & 1 & von Essen \etal (2014) \\
1388 & 2455856.40365$^{\rm{b}}$ & 0.00070$^{\rm{b}}$ & 0.00075$^{\rm{b}}$ & 1 & von Essen \etal (2014) \\
1389 & 2455857.62609$^{\rm{b}}$ & 0.00080$^{\rm{b}}$ & 0.00081$^{\rm{b}}$ & 1 & von Essen \etal (2014) \\
1398 & 2455868.61036$^{\rm{b}}$ & 0.00071$^{\rm{b}}$ & 0.00067$^{\rm{b}}$ & 1 & von Essen \etal (2014) \\
1415 & 2455889.33922$^{\rm{b}}$ & 0.00097$^{\rm{b}}$ & 0.00098$^{\rm{b}}$ & 1 & von Essen \etal (2014) \\
1447 & 2455928.37440$^{\rm{b}}$ & 0.00065$^{\rm{b}}$ & 0.00065$^{\rm{b}}$ & 1 & von Essen \etal (2014) \\
1456 & 2455939.35426$^{\rm{b}}$ & 0.00046$^{\rm{b}}$ & 0.00047$^{\rm{b}}$ & 1 & von Essen \etal (2014) \\
1526 & 2456024.74734$^{\rm{b}}$ & 0.00014$^{\rm{b}}$ & 0.00014$^{\rm{b}}$ & 1 & Zhang \etal (2018) \\
1530 & 2456029.62679$^{\rm{b}}$ & 0.00016$^{\rm{b}}$ & 0.00016$^{\rm{b}}$ & 1 & Zhang \etal (2018) \\
1639 & 2456162.59906$^{\rm{b}}$ & 0.0013$^{\rm{b}}$ & 0.0014$^{\rm{b}}$ & 1 & von Essen \etal (2014) \\
1680 & 2456212.60893$^{\rm{b}}$ & 0.00033$^{\rm{b}}$ & 0.00031$^{\rm{b}}$ & 1 & von Essen \etal (2014) \\
1684 & 2456217.48687$^{\rm{b}}$ & 0.00049$^{\rm{b}}$ & 0.00048$^{\rm{b}}$ & 1 & von Essen \etal (2014) \\
2226 & 2456878.65777 & 0.00033 & 0.00032 & 1 & this paper \\
2244 & 2456900.61530 & 0.00036 & 0.00035 & 1 & this paper \\
2272 & 2456934.77146$^{\rm{b}}$ & 0.00059$^{\rm{b}}$ & 0.00059$^{\rm{b}}$ & 1 & Johnson \etal (2015) \\
2943 & 2457753.30433 & 0.00052 & 0.00053 & 1 & this paper \\
2952 & 2457764.28369 & 0.00043 & 0.00041 & 1 & this paper \\
3158 & 2458015.57583 & 0.00046 & 0.00047 & 1 & this paper \\
3167 & 2458026.55466 & 0.00077 & 0.00078 & 1 & this paper \\
3207 & 2458075.35041 & 0.00037 & 0.00037 & 1 & this paper \\
3458 & 2458381.53678 & 0.00055 & 0.00054 & 1 & this paper \\
3476 & 2458403.49659 & 0.00045 & 0.00046 & 1 & this paper \\
3498 & 2458430.33394 & 0.00056 & 0.00058 & 1 & this paper \\
3503 & 2458436.43219 & 0.00034 & 0.00034 & 1 & this paper \\
\hline
\multicolumn{6}{l}{$N_{\rm{tr}}$ is the number of individual light curves taken to determine $T_{\rm{mid}}$.} \\
\multicolumn{6}{l}{$^{\rm{a}}$ Epoch 0 set for $T_0$ from Collier Cameron \etal (2010).}\\
\multicolumn{6}{l}{$^{\rm{b}}$ Value taken from the original paper.}\\
}

\MakeTable{ r l l l c l}{12.5cm}{New and redetermined literature mid-transit times for WASP-103 b.}
{\hline
 Epoch$^{\rm{a}}$ & $T_{\rm{mid}}$ (BJD$_{\rm{TDB}}$) & \multicolumn{1}{c}{$+\sigma$ (d)} & \multicolumn{1}{c}{$-\sigma$ (d)} & $N_{\rm{tr}}$ & Light curve source\\
\hline
   0 & 2456459.59968 & 0.00087 & 0.00095 & 1 & Gillon \etal (2014) \\
  14 & 2456472.55818 & 0.00095 & 0.00096 & 1 & Gillon \etal (2014) \\
  28 & 2456485.51480 & 0.00086 & 0.00088 & 1 & Gillon \etal (2014) \\
  41 & 2456497.54630 & 0.00076 & 0.00075 & 1 & Gillon \etal (2014) \\
  54 & 2456509.57806 & 0.00070 & 0.00068 & 1 & Gillon \etal (2014) \\
 333 & 2456767.80583 & 0.00028 & 0.00027 & 1 & Southworth \etal (2014) \\
 346 & 2456779.83874 & 0.00040 & 0.00040 & 1 & Southworth \etal (2014) \\
 387 & 2456817.78569 & 0.00036 & 0.00037 & 1 & Southworth \etal (2014) \\
 389 & 2456819.63655 & 0.00051 & 0.00053 & 1 & Delrez \etal (2017) \\
 402 & 2456831.66832 & 0.00046 & 0.00047 & 1 & Southworth \etal (2014) \\
 403 & 2456832.59399 & 0.00036 & 0.00039 & 1 & Southworth \etal (2014) \\
 416 & 2456844.62624 & 0.00017 & 0.00017 & 6 & Southworth \etal (2014) \\
  & & & & & \& Delrez \etal (2017) \\
 429 & 2456856.65842 & 0.00025 & 0.00024 & 1 & Southworth \etal (2014) \\
 430 & 2456857.58397 & 0.00012 & 0.00011 & 5 & Southworth \etal (2014) \\
 457 & 2456882.57428 & 0.00053 & 0.00054 & 1 & Delrez \etal (2017) \\
 776 & 2457177.82206 & 0.00104 & 0.00100 & 1 & Turner \etal (2017) \\
 801 & 2457200.96071 & 0.00037 & 0.00036 & 1 & Lendl \etal (2017) \\
 815 & 2457213.91777 & 0.00034 & 0.00034 & 1 & Lendl \etal (2017) \\
1137 & 2457511.94421 & 0.00010 & 0.00011 & 1 & Lendl \etal (2017) \\
1142 & 2457516.57160 & 0.00075 & 0.00075 & 1 & this paper \\
1143 & 2457517.49799 & 0.00104 & 0.00116 & 1 & this paper \\
1144 & 2457518.42426 & 0.00101 & 0.00095 & 1 & this paper \\
1532 & 2457877.53468 & 0.00063 & 0.00064 & 1 & this paper \\
1546 & 2457890.49304 & 0.00058 & 0.00060 & 1 & this paper \\
1586 & 2457927.51425 & 0.00040 & 0.00039 & 1 & this paper \\
1934 & 2458249.60409 & 0.00029 & 0.00029 & 1 & this paper \\
1935 & 2458250.52991 & 0.00031 & 0.00031 & 1 & this paper \\
1962 & 2458275.51920 & 0.00028 & 0.00027 & 2 & this paper \\
1976 & 2458288.47740 & 0.00030 & 0.00029 & 1 & this paper \\
1977 & 2458289.40295 & 0.00055 & 0.00055 & 1 & this paper \\
2032 & 2458340.30742 & 0.00047 & 0.00045 & 1 & this paper \\
\hline
\multicolumn{6}{l}{$N_{\rm{tr}}$ is the number of individual light curves taken to determine $T_{\rm{mid}}$.} \\
\multicolumn{6}{l}{$^{\rm{a}}$ Epoch 0 set for $T_0$ from Gillon \etal (2014).}\\
}

\end{document}